\documentclass{aa}
\usepackage{graphicx}
\usepackage{txfonts}
\usepackage{subfigure}
\usepackage{multirow}
\usepackage[colorlinks=true, linkcolor=blue, urlcolor=blue, citecolor=blue]{hyperref}

\begin{document}

\title{Recovering the pattern speeds of edge-on barred galaxies via an orbit-superposition method}
\titlerunning{Recovering the pattern speeds of edge-on barred galaxies}

\author
{Yunpeng Jin\inst{1}\thanks{E-mail: jyp199333@163.com}
\and Ling Zhu\inst{2}\thanks{Corr author: lzhu@shao.ac.cn}
\and Behzad Tahmasebzadeh\inst{3}
\and Shude Mao\inst{1}
\and Glenn van de Ven\inst{4}
\and Rui Guo\inst{5}
\and Runsheng Cai\inst{2}}

\institute
{Department of Astronomy, Westlake University, Hangzhou, Zhejiang 310030, China
\and Shanghai Astronomical Observatory, Chinese Academy of Sciences, 80 Nandan Road, Shanghai 200030, China
\and Department of Astronomy, University of Michigan, Ann Arbor, MI, 48109, USA
\and Department of Astrophysics, University of Vienna, Türkenschanzstraße 17, 1180 Wien, Austria
\and National Astronomical Observatories, Chinese Academy of Sciences, 20A Datun Road, Chaoyang District, Beĳing 100101, China}
\date{}

\abstract
{We developed an orbit-superposition method for edge-on barred galaxies and evaluated its capability to recover the bar pattern speed $\rm\Omega_p$. We selected three simulated galaxies (Au-18, Au-23, and Au-28) with known pattern speeds from the Auriga simulations and created MUSE-like mock data sets with edge-on views (inclination angles $\theta_{\rm T}\ge85^\circ$) and various bar azimuthal angles $\varphi_{\rm T}$. For mock data sets with side-on bars ($\varphi_{\rm T}\ge50^\circ$), the model-recovered pattern speeds $\rm\Omega_p$ encompass the true pattern speeds $\rm\Omega_T$ within the model uncertainties ($1\sigma$ confidence levels, $68\%$) for 10 of 12 cases. The average model uncertainty within the $1\sigma$ confidence levels is equal to $10\%$. For mock data sets with end-on bars ($\varphi_{\rm T}\le30^\circ$), the model uncertainties of $\rm\Omega_p$ depend significantly on the bar azimuthal angles $\varphi_{\rm T}$, with the uncertainties of cases with $\varphi_{\rm T}=10^\circ$ approaching $\sim30\%$. However, by imposing a stricter constraint on the bar morphology ($p_{\rm bar}\le0.50$), the average uncertainties are reduced to $14\%$ , and $\rm\Omega_p$ still encompass $\rm\Omega_T$ within the model uncertainties for three of four cases. For all the models that we create in this paper, the $2\sigma$ ($95\%$) confidence levels of the model-recovered pattern speeds $\rm\Omega_p$ always cover the true values $\rm\Omega_T$.}


\maketitle
\begin{nolinenumbers}

\section{Introduction}
In recent years, many photometric studies from optical to near-infrared wavelengths have shown that bars exist in more than half of the nearby disk galaxies ($z\lesssim 0.1$; e.g. \citealp{Eskridge2000,Knapen2000,Marinova2007,MenendezDelmestre2007,Barazza2008,Aguerri2009,Masters2011,Buta2015,Erwin2018}). Bars play an important role in galaxy formation and evolution by redistributing the angular momenta of their host galaxies, resulting in the gas inflow from disks to their centres (e.g. \citealp{Athanassoula2003,Kormendy2004,Gadotti2011}). The bars are usually described by three main properties: bar length, bar strength, and pattern speed $\rm\Omega_p$, which is defined as the angular rotational frequency of the bar. The bar length and bar strength can be estimated directly from galaxy images, using methods such as isophote fitting (e.g. \citealp{Marinova2007,Aguerri2009}) and structural decomposition of surface brightness (e.g. \citealp{Laurikainen2005,Gadotti2008,Gadotti2011}). However, estimating the bar pattern speed $\rm\Omega_p$ requires kinematics extracted from spectroscopy.

In the past decades, many attempts have been made to estimate $\rm\Omega_p$, such as inferring $\rm\Omega_p$ by matching the observed gas distributions and/or gas velocities with those of simulated galaxies (e.g. \citealp{Weiner2001,Rautiainen2008,Treuthardt2008}) and deriving $\rm\Omega_p$ based on galaxy morphology features correlated with Lindblad resonances (e.g. \citealp{Buta1986,Athanassoula1992,Canzian1993,Aguerri1998,Buta2009}). However, these methods are indirect or strongly dependent on the model description of morphological features. A model-independent method for measuring $\rm\Omega_p$ based on long-slit spectroscopic data is the \citet{Tremaine1984} method (TW method). This method has been widely used to study the pattern speeds of individual galaxies (e.g. \citealp{Kent1987,Merrifield1995,Gerssen1999,Debattista2002,Debattista2004,Aguerri2003,Corsini2003,Zimmer2004}).

In recent years, integral field spectroscopy (IFS) surveys such as SAURON \citep{Bacon2001}, $\rm ATLAS^{3D}$ \citep{Cappellari2011}, CALIFA \citep{Sanchez2012}, SAMI \citep{Bryant2015}, MaNGA \citep{Bundy2015}, and multiple VLT/MUSE programmes \citep{Bacon2017} have observed thousands of nearby galaxies. IFS surveys provide the spectra over 2D fields of galaxies, and thus the stellar kinematic maps of these galaxies can be derived. The TW method has been updated and applied in IFS surveys to measure $\rm\Omega_p$ of $\sim$100 galaxies (e.g. \citealp{Aguerri2015,Cuomo2019,Guo2019,GarmaOehmichen2020,GarmaOehmichen2022}), but tests with mock IFS data show that accurate estimations of $\rm\Omega_p$ are only obtained for limited inclination angles and bar angles: The galaxy needs to be observed from a non-edge-on view (inclination angle $\theta\lesssim70^\circ$, where $\theta=90^\circ$ implies perfectly edge-on), and the bar cannot be too parallel nor too perpendicular to the disk major axis (e.g. \citealp{Guo2019,Zou2019}). This indicates that the TW method is not able to measure the pattern speeds of edge-on barred galaxies.

On the other hand, IFS data provide an opportunity to estimate the bar pattern speed $\rm\Omega_p$ through dynamical modelling. The Schwarzschild's orbit-superposition method \citep{Schwarzschild1979,Schwarzschild1982,Schwarzschild1993} is a flexible technique for building dynamical models of galaxies by fitting the luminosity distributions and stellar kinematic maps. Several implementations of Schwarzschild’s method are currently in use (e.g. \citealp{vdB2008,Long2018,Vasiliev2020,Neureiter2021,Quenneville2022}). The \citet{vdB2008} implementation along with its publicly available version DYNAMITE \citep{Jethwa2020}\footnote{\url{https://dynamics.univie.ac.at/dynamite_docs/}} and its modified version TriOS \citep{Quenneville2022} were extensively used to analyse the properties of galaxies in IFS surveys, including the mass distributions and intrinsic shapes (e.g. \citealp{Jin2019,Jin2020,Santucci2022,Santucci2024,Pilawa2022,Pilawa2024,Thater2023}), internal orbital structures (e.g. \citealp{Zhu2018a,Zhu2018b,Zhu2018c,Jin2019,Jin2020,Santucci2022,Santucci2023,Santucci2024,Thater2023}), and black hole masses (e.g. \citealp{Quenneville2021,Quenneville2022,Pilawa2022,Pilawa2024,Liepold2025}). This implementation was further updated to support the modelling of stellar populations (e.g. \citealp{Poci2019,Poci2021,Zhu2020,Zhu2022a,Zhu2022b,Ding2023,Jin2024}) and gas kinematics (e.g. \citealp{Yang2020,Yang2024}). Recently, an orbit-superposition method for barred galaxies based on this implementation was developed (\citealp{Tahmasebzadeh2021,Tahmasebzadeh2022}; see also the public code DYNAMITE). Using MUSE-like mock data, this method was able to recover $\rm\Omega_p$ with an accuracy of $\sim10\%$ in the cases where the galaxy is non-edge-on ($\theta\lesssim80^\circ$) and the bar is explicitly parallel or perpendicular to the disk major axis \citep{Tahmasebzadeh2022}, and was applied to a barred S0 galaxy NGC 4371 \citep{Tahmasebzadeh2024}.

For barred galaxies, edge-on observations provide information that cannot be observed from face-on views, such as the boxy, peanut, or X-shaped (hereafter BP/X-shaped) structure, which is considered to be caused by the bar buckling (e.g. \citealp{Raha1991,Kuijken1995,Athanassoula2005,Debattista2006,MartinezValpuesta2006}). Recently, an ongoing VLT/MUSE programme, GECKOS \citep{vdSande2024}, came out. GECKOS is designed to observe 36 Milky Way-like edge-on ($\theta\gtrsim85^\circ$) galaxies and will provide high-quality IFS data of these galaxies. Eight of the 12 galaxies in the first GECKOS internal data release are barred and have BP/X-shaped structures \citep{FraserMcKelvie2024}. Estimating bar pattern speed $\rm\Omega_p$ for edge-on barred galaxies was rarely considered in previous studies. \citet{Dattathri2024} constructed orbit-superposition dynamical modelling for a BP/X-shaped barred galaxy from an N-body simulation using the \citet{Vasiliev2020} implementation FORSTAND and recovered $\rm\Omega_p$ with an accuracy of $\sim10\%$. However, their results rely on the parametric functions for BP/X-shaped structures, where some parameters (including the viewing angles) are fixed during the fitting process.

In this paper, we expand the barred orbit-superposition method developed by \citet{Tahmasebzadeh2021,Tahmasebzadeh2022} to edge-on galaxies, with one key objective being to measure the bar pattern speeds of these galaxies. We validate the method using simulated galaxies in the Auriga simulations \citep{Grand2017,Grand2024}, allowing the intrinsic 3D density and viewing angles to be determined by the modelling. This means that our method can be applied to real observations such as GECKOS to derive $\rm\Omega_p$ in the future.

The structure of the paper is as follows. In Sect.~\ref{sec2}, we create mock data using simulated galaxies. In Sect.~\ref{sec3}, we introduce a methodology for creating a barred orbit-superposition model from mock data with a set of free parameters. In Sect.~\ref{sec4}, we search for the best-fitting model in the parameter space and compare it with the truth. In Sect.~\ref{sec5}, we present the recovery of $\rm\Omega_p$. We discuss our results in Sect.~\ref{sec6} and summarise in Sect.~\ref{sec7}. The cosmological parameters adopted in this paper are $H_{\rm 0}=\rm 67.4\,km\cdot s^{-1}\cdot Mpc^{-1}$, $\Omega_{\Lambda}=0.685$, and $\Omega_{\rm m}=0.315$ \citep{PlanckCollaboration2020}. Throughout the paper, the galaxy's intrinsic coordinate system is labelled as $(x,y,z)$, where the $x$-axis aligns with the bar major axis and the $z$-axis is perpendicular to the disk plane to distinguish it from the coordinates in the observing plane $(x',y')$.

\section{Simulated galaxies and mock data}
\label{sec2}
\subsection{The Auriga simulations}
\label{sec2.1}
\begin{figure*}
    \centering
    \includegraphics[width=14cm]{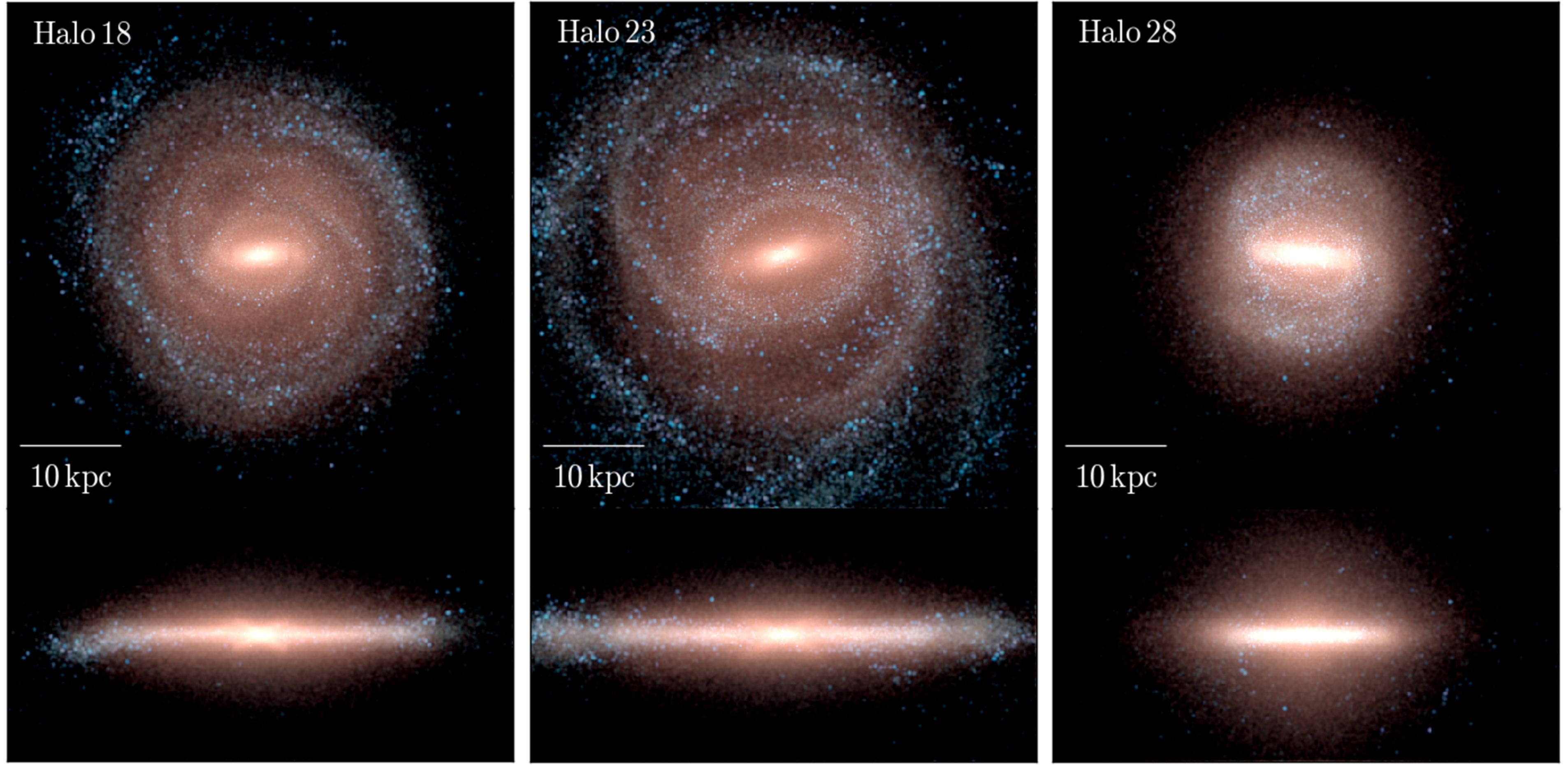}
    \caption{Face-on and edge-on images of the three Auriga simulated galaxies, Au-18, Au-23, and Au-28. The images come from \url{https://wwwmpa.mpa-garching.mpg.de/auriga/data}.}
    \label{image}
\end{figure*}

The Auriga project is a suite of cosmological magnetohydrodynamical zoom-in simulations that consists of 30 isolated Milky Way-mass halos \citep{Grand2017}. The Auriga sample was selected from the dark matter-only version of the EAGLE Ref-L100N1504 simulation \citep{Schaye2015}, and the simulations were carried out with the N-body magnetohydrodynamical moving-mesh code AREPO \citep{Springel2010}. The simulations contain comprehensive physical processes of galaxy formation, such as primordial and metal-line cooling mechanisms \citep{Vogelsberger2013}, a hybrid multi-phase star formation model \citep{Springel2003}, stellar feedback \citep{Marinacci2014}, active galactic nucleus feedback \citep{Springel2005}, an ultraviolet background field \citep{FaucherGiguere2009}, and magnetic fields \citep{Pakmor2013}. The initial masses of baryon particles and high-resolution dark matter particles were $m_{\rm b}\sim5\times10^4\,\rm M_\odot$ and $m_{\rm DM}\sim3\times10^5\,\rm M_\odot$, respectively. We refer to \citet{Grand2017} for more details on the Auriga simulations.

\subsection{Sample galaxies and mock data}
\label{sec2.2}
\begin{figure*}
    \centering
    \includegraphics[width=12cm]{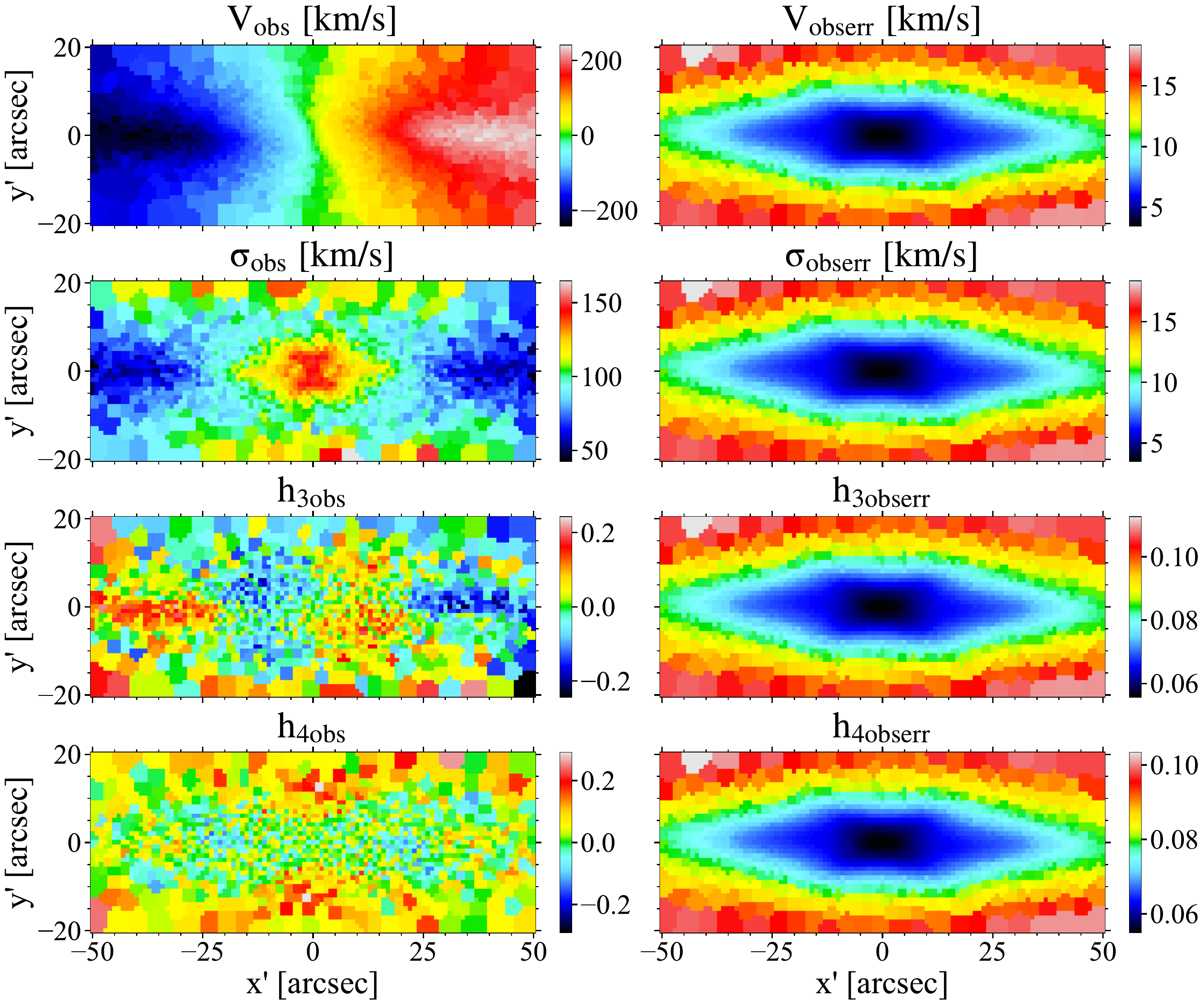}
    \caption{Mock kinematic maps and error maps for Au-23 with viewing angles $(\theta_{\rm T},\varphi_{\rm T})=(85^\circ,50^\circ)$ (denoted as Au-23-85-50 hereafter). Left panels represent the mock kinematic maps and the right panels represent the corresponding error maps. From top to bottom: Mean velocity $V$, velocity dispersion $\sigma$, third-order Gauss-Hermite coefficient $h_3$, and fourth-order Gauss-Hermite coefficient $h_4$. We note that the error maps are generated from particle noise, and therefore exhibit similar statistical properties, but their ranges are different (as indicated by the colour bars in the right panels).}
    \label{kinematic-data}
\end{figure*}

\begin{table*}
\centering
\caption{Galaxy properties from \citet{Grand2017} and our mock data sets for Au-18, Au-23, and Au-28.}
\begin{tabular}{|c|c|c|c|c|c|c|}
\hline
\multicolumn{4}{|c|}{Galaxy properties} & \multicolumn{3}{|c|}{Mock data sets}\\
\hline
Name & $M_*\ [\rm M_\odot]$ & $M_{200}\ [\rm M_\odot]$ & $R_{\rm opt}$ [kpc] & $(\theta_{\rm T},\varphi_{\rm T})$ & Classification & BP/X-shaped structure\\
\hline
\multirow{4}*{Au-18} & \multirow{4}*{$8.04\times10^{10}$} & \multirow{4}*{$1.22\times10^{12}$} & \multirow{4}*{21} & $(85^\circ,50^\circ)$ & \multirow{4}*{side-on bar} & \multirow{8}*{Yes} \\
~ & ~ & ~ & ~ & $(85^\circ,80^\circ)$ & ~ & ~ \\
~ & ~ & ~ & ~ & $(89^\circ,50^\circ)$ & ~ & ~ \\
~ & ~ & ~ & ~ & $(89^\circ,80^\circ)$ & ~ & ~ \\
\cline{1-6}
\multirow{8}*{Au-23} & \multirow{8}*{$9.02\times10^{10}$} & \multirow{8}*{$1.58\times10^{12}$} & \multirow{8}*{25} & $(85^\circ,50^\circ)$ & \multirow{4}*{side-on bar} & ~ \\
~ & ~ & ~ & ~ & $(85^\circ,80^\circ)$ & ~ & ~ \\
~ & ~ & ~ & ~ & $(89^\circ,50^\circ)$ & ~ & ~ \\
~ & ~ & ~ & ~ & $(89^\circ,80^\circ)$ & ~ & ~ \\
\cline{5-7}
~ & ~ & ~ & ~ & $(85^\circ,10^\circ)$ & \multirow{4}*{end-on bar} & \multirow{8}*{No} \\
~ & ~ & ~ & ~ & $(85^\circ,30^\circ)$ & ~ & ~ \\
~ & ~ & ~ & ~ & $(89^\circ,10^\circ)$ & ~ & ~ \\
~ & ~ & ~ & ~ & $(89^\circ,30^\circ)$ & ~ & ~ \\
\cline{1-6}
\multirow{4}*{Au-28} & \multirow{4}*{$10.45\times10^{10}$} & \multirow{4}*{$1.61\times10^{12}$} & \multirow{4}*{17.5} & $(85^\circ,50^\circ)$ & \multirow{4}*{side-on bar} & ~ \\
~ & ~ & ~ & ~ & $(85^\circ,80^\circ)$ & ~ & ~ \\
~ & ~ & ~ & ~ & $(89^\circ,50^\circ)$ & ~ & ~ \\
~ & ~ & ~ & ~ & $(89^\circ,80^\circ)$ & ~ & ~ \\
\hline
\end{tabular}
\tablefoot{From left to right: (1) Galaxy name; (2) total stellar mass $M_*$; (3) virial mass of dark matter halo $M_{\rm 200}$; (4) optical radius $R_{\rm opt}$; (5) inclination angle $\theta_{\rm T}$ (from face-on $0^\circ$ to edge-on $90^\circ$) and bar azimuthal angle $\varphi_{\rm T}$ (from end-on $0^\circ$ to side-on $90^\circ$); (6) classification of the projected bar; (7) whether BP/X-shaped structure is considered in the modelling. We note that $\varphi_{\rm T}=0^\circ$ indicates the bar is perfectly end-on while $\varphi_{\rm T}=90^\circ$ means the bar is perfectly side-on.}
\label{table-all-angles}
\end{table*}

We selected three simulated galaxies with strong bars from the Auriga simulations: Au-18, Au-23, and Au-28. Au-18 and Au-23 have clear BP/X-shaped structures, while Au-28 does not. Their stellar masses are $M_*=8.04\times10^{10}\,\rm M_\odot$, $9.02\times10^{10}\,\rm M_\odot$, and $10.45\times10^{10}\,\rm M_\odot$, respectively, and their corresponding dark matter halo masses are $M_{\rm 200}=1.22\times10^{12}\,\rm M_\odot$, $1.58\times10^{12}\,\rm M_\odot$, and $1.61\times10^{12}\,\rm M_\odot$. The face-on and edge-on images of these three galaxies are shown in Fig.~\ref{image}.

We took snapshots at $z=0$ and placed the galaxies at a distance of $\rm41.25\,Mpc$ ($\rm 1\,arcsec=0.2\,kpc$). Then we created our mock data sets by projecting the stellar particles onto observing planes with various inclination angles $\theta_{\rm T}$ and bar azimuthal angles $\varphi_{\rm T}$ (true values are labelled with the subscript `T'). We note that $\varphi_{\rm T}=0^\circ$ indicates the bar is perfectly end-on, and $\varphi_{\rm T}=90^\circ$ means the bar is perfectly side-on. For each galaxy, we chose four projections with side-on bars ($\varphi_{\rm T}\ge50^\circ$). For Au-23, we further chose four projections with end-on bars ($\varphi_{\rm T}\le30^\circ$). For each projection, we created a mock data set, and therefore we have $3\times4+4=16$ mock data sets in total. Table~\ref{table-all-angles} lists the basic properties of these three simulated galaxies and all projections we use. Each mock data set is labelled with the format `name-$\theta_{\rm T}$-$\varphi_{\rm T}$', for example, the mock observation of Au-23 with viewing angles $(\theta_{\rm T},\varphi_{\rm T})=(85^\circ,50^\circ)$ is denoted as Au-23-85-50. This notation is used consistently in the paper. We take Au-23-85-50 as an example to demonstrate the mock data and the model results. This example is classified as a side-on case, with the BP/X-shaped structure clearly visible in the projected image.

We constructed the mock surface brightness from the stellar mass distribution by assuming a constant stellar mass-to-light ratio, $M_{\star}/L=2$, and a pixel size of $\rm1\times1\,arcsec^2$. For each pixel, we calculated the signal-to-noise ratio ($S/N$) by taking the particle numbers as the signal and assuming Poisson noise. We performed the Voronoi binning method \citep{Cappellari2003} to obtain apertures with sizes similar to those of MUSE/VLT observed real galaxies, which is accomplished by setting a target $S/N=30$. Then we derived the kinematic maps by fitting the velocity distribution of stellar particles in each binned aperture with the Gauss-Hermite coefficients, including the luminosity-weighted mean velocity $V$, the velocity dispersion $\sigma$, and the higher-order coefficients $h_3$ and $h_4$. After that, we perturbed the kinematic data in each aperture by adding Gaussian noise following the method in \citet{Tsatsi2015}. Finally, we point-symmetrised the kinematic data in the apertures. The kinematic map and the corresponding error map for Au-23-85-50 are shown in Fig.~\ref{kinematic-data}. The kinematic maps together with the surface brightness are taken as the observational constraints for the barred orbit-superposition method.

\section{Creating a barred orbit-superposition model}
\label{sec3}
The steps in creating an orbit-superposition model for a galaxy include \citep{vdB2008}: (1) constructing the gravitational potential using a set of free parameters, (2) sampling and integrating orbits, and (3) solving orbit weights by fitting the observational data. When modelling a non-edge-on barred galaxy, the construction of gravitational potential was modified to incorporate a rotating bar \citep{Tahmasebzadeh2021,Tahmasebzadeh2022}. In this work, we further update the construction of potential to support edge-on barred galaxies with possible BP/X-shaped structures. We demonstrate these three steps in Sects.~\ref{sec3.1},~\ref{sec3.2}, and~\ref{sec3.3} using the example case Au-23-85-50.

\subsection{Gravitational potential}
\label{sec3.1}
\begin{figure}
    \centering
    \includegraphics[width=8.5cm]{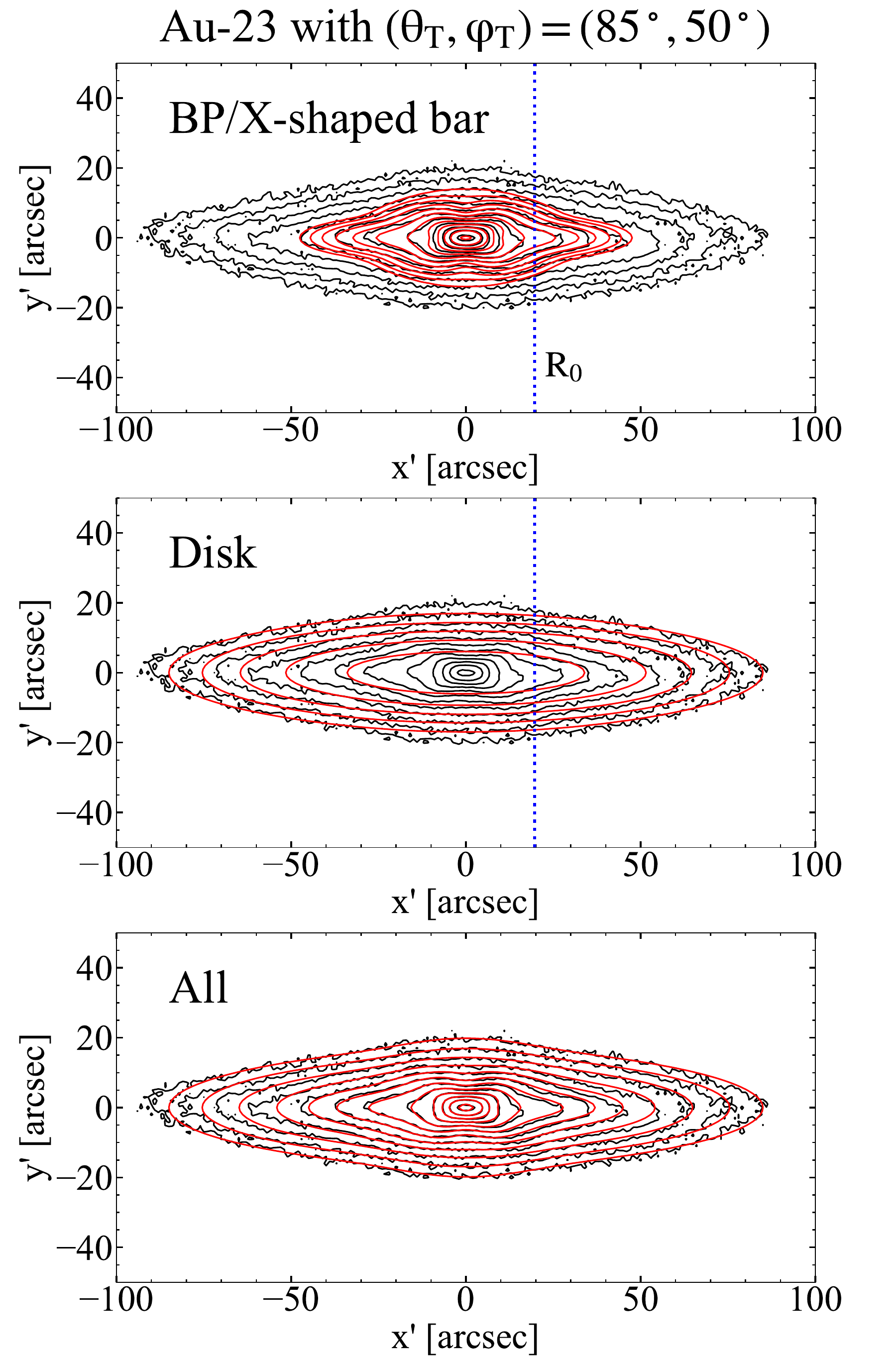}
    \caption{True surface brightness contours and MGE fitting results in the observing plane $x'$-$y'$ for Au-23-85-50. The black contours in all panels represent the true surface brightness. The red contours in the top, middle and bottom panel denote the bar component with BP/X-shaped structure, the disk component and all components in the MGE fitting results, respectively. The contour interval is equal to 0.5 magnitude. The blue dashed line indicates the separating radius $R_0$ that distinguishes Gaussians as either part of the bar component or the disk component in the MGE fitting process (see Eq.~\ref{R0}).}
    \label{mge-fitting}
\end{figure}
\begin{figure*}
    \centering
    \includegraphics[width=16cm]{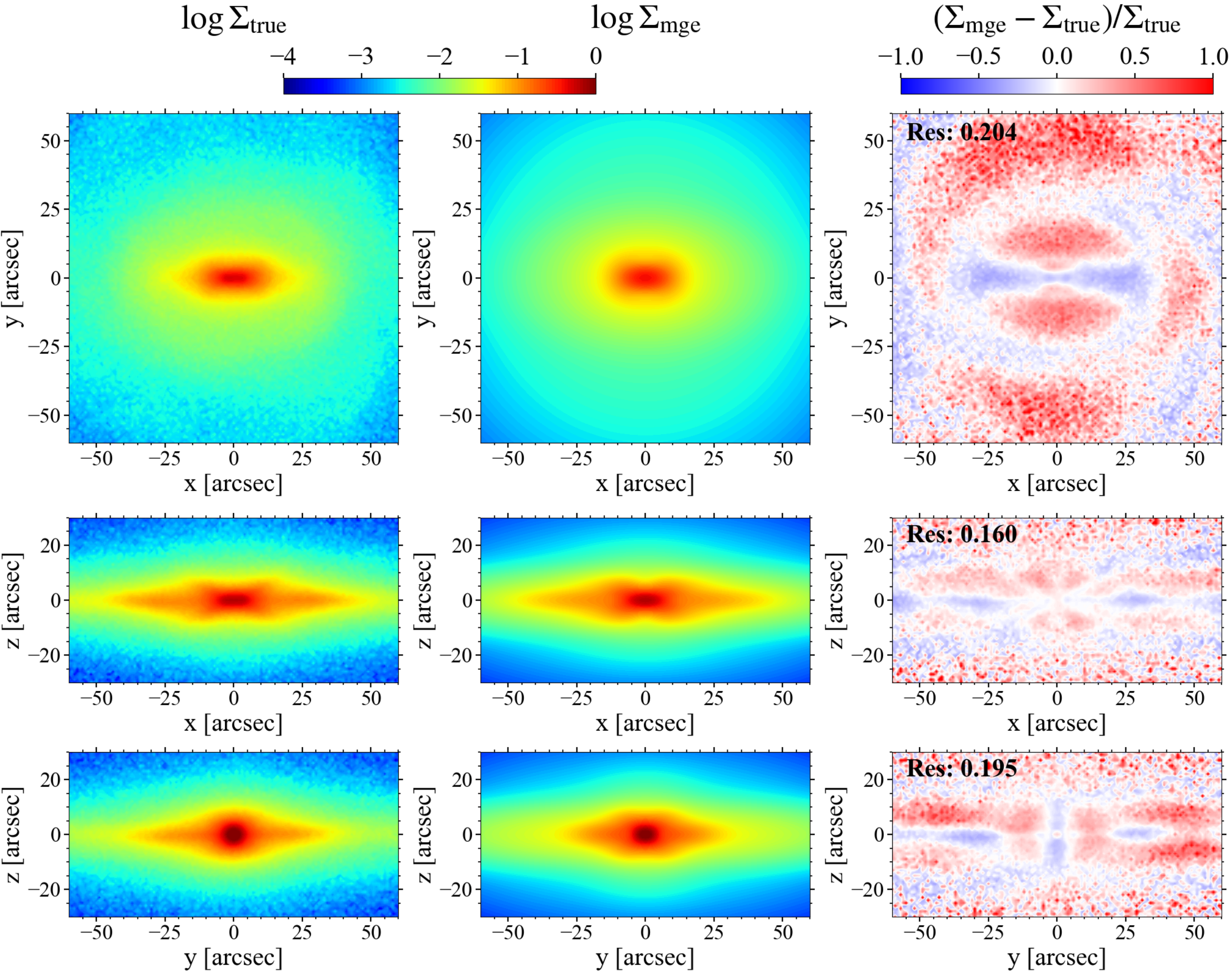}
    \caption{Comparison of true luminosity distribution and MGE-predicted luminosity distribution for Au-23-85-50. The true luminosity distribution is derived from the stellar particles in simulations by assuming a constant mass-to-light ratio $(M_{\star}/L)_{\rm T}=2$. Left panels: from top to bottom, the panels show the normalised true luminosity distribution $\Sigma_{\rm true}$ of Au-23 in the intrinsic $x$-$y$, $x$-$z$, and $y$-$z$ planes. Middle panels: the corresponding MGE-predicted luminosity distribution $\Sigma_{\rm mge}$, which is deprojected from 2D MGE under the true viewing angles $(\theta,\varphi)=(\theta_{\rm T},\varphi_{\rm T})=(85^\circ,50^\circ)$. Right panels: the residuals between the true and the MGE-predicted luminosity distribution $(\Sigma_{\rm true}-\Sigma_{\rm mge})/\Sigma_{\rm true}$. The value in each panel represents the average absolute value of the residuals.}
    \label{mge-deprojection}
\end{figure*}

The gravitational potential of the galaxy is generated by a combination of stellar mass and dark matter mass. We do not include a central black hole, as the spatial resolution of kinematic data we use in this paper ($\rm\sim0.2\,kpc/pixel$) is not high enough to accurately determine the black hole mass. Therefore, including a black hole in the modelling does not significantly influence our results.

\subsubsection{Stellar potential}
\label{sec3.1.1}
In the modelling, the stellar potential combines contributions from a triaxial bar and an axisymmetric disk, which are both constructed by deprojecting the galaxy's surface brightness. The construction procedure involves five steps: (1) determining a separating radius $R_0$ between the bar and the disk, (2) fitting the surface brightness of the entire galaxy with the multi-Gaussian expansion (MGE) formalism (\citealp{Emsellem1994,Cappellari2002}) and assigning Gaussian components to the bar or the disk by comparing their scales to $R_0$, (3) deprojecting the 2D Gaussians of the bar and the disk into 3D individually under specified viewing angles and then summing their contributions to derive the 3D luminosity distribution, (4) multiplying the 3D luminosity density by a constant mass-to-light ratio $M_{\star}/L$, and (5) deriving the stellar potential by solving Poisson's equation.

To determine the separating radius $R_0$, the galaxy's surface brightness is first modelled by a combination of a S\'ersic bar and an exponential disk using the structural decomposition algorithm GALFIT \citep{Peng2002,Peng2010}. This yields the effective radius of the S\'ersic bar $r_e$ and the scale length of the exponential disk $r_s$. The separating radius $R_0$ along the disk major axis is then defined as
\begin{equation}
    R_0= \min (3r_e, r_s).
\label{R0}
\end{equation}
This radius $R_0$ is employed to distinguish Gaussians in the MGE fitting process as either part of the triaxial bar or the axisymmetric disk.

The surface brightness of the entire galaxy is then fitted by the MGE formalism. From edge-on views, the misalignment between the bar major axis and the disk major axis is usually small and is difficult to measure accurately. To simplify the MGE fitting process, we restrict all Gaussians to have the same major axis that aligns with the $x'$-axis. Therefore, the surface brightness fitted by MGE can be expressed as \citep{Cappellari2002}
\begin{equation}
    \Sigma(x',y')=\sum_{j=1}^{N} \frac{L_j}{2\pi\sigma_j'^2 q_j'}\exp \left[-\frac{1}{2\sigma_j'^2} \left(x'^2+\frac{y'^2}{q_j'^2} \right) \right],
\label{mge2D}
\end{equation}
where $N$ is the number of Gaussians, the subscript $j$ corresponds to the $j$-th Gaussian, $L_j$ is the total luminosity, $q_j'=b_j'/a_j'$ is the axis ratio and $\sigma_j'$ is the dispersion along the major axis $x'$. The Gaussians with dispersion $\sigma_j'\le R_0$ are taken as the bar component, while the Gaussians with $\sigma_j'>R_0$ are taken as the disk component. If the BP/X-shaped structure exists in the galaxy image, the Gaussians of the bar component are allowed to have negative luminosity ($L_j<0$), in order to fit the BP/X-shaped structure. The MGE fitting parameters for the surface brightness of Au-23-85-50 are presented in Table~\ref{table-mge-parameters}, with the isophotal contours plotted in Fig.~\ref{mge-fitting}. The BP/X-shaped structure in the galaxy's central region is successfully fitted.

During the deprojection process, the morphology of the triaxial bar depends on its inclination angle $\theta_{\rm bar}$, intrinsic azimuthal angle $\varphi$, and position angle $\psi_{\rm bar}$ in the observing plane, while that of the axisymmetric disk is affected by its inclination angle $\theta_{\rm disk}$ and position angle $\psi_{\rm disk}$. Under the fundamental assumption that the bar major axis ($x$-axis) aligns with the disk major axis, the inclination angles of both components become equal ($\theta_{\rm bar}=\theta_{\rm disk}=\theta$), and the disk major axis in the observing plane becomes aligned with $x'$-axis ($\psi_{\rm disk}=90^\circ$).

Therefore, for the axisymmetric disk, only the inclination angle $\theta$ is free. Each 2D Gaussian belonging to the disk can be deprojected to a 3D Gaussian by
\begin{equation}
    q_j^2= \frac{q_j'^2-\cos^2\theta}{\sin^2\theta},
\end{equation}
and
\begin{equation}
    \sigma_j'=\sigma_j,
\end{equation}
where $q_j=c_j/a_j$ represents the short-to-major axis ratio of the $j$-th 3D Gaussian ($a_j\ge b_j\ge c_j$), and $\sigma_j$ is the dispersion along the major axis $x$. By combining all 3D Gaussians belonging to the disk, we transform the inclination angle $\theta$ into the disk axis ratio $q_{\rm disk}$, which is defined as the axis ratio of the disk isophote at $x=\rm60\,arcsec$ in the intrinsic $x$-$z$ plane. This ratio serves as an equivalent free parameter instead of $\theta$ in the modelling.

For the triaxial bar, each 3D Gaussian can be calculated from the viewing angles $(\theta,\varphi,\psi_{\rm bar})$ by (\citealp{Cappellari2002,vdB2008})
\begin{equation}
    1-q_j^2=\frac{\delta_j'[2\cos2\psi+\sin2\psi(\sec\theta\cot\varphi-\cos\theta\tan\varphi)]}{2\sin^2\theta[\delta_j'\cos\psi(\cos\psi+\cot\varphi\sec\theta\sin\psi)-1]},
\label{pqu1}
\end{equation}
\begin{equation}
    p_j^2-q_j^2=\frac{\delta_j'[2\cos2\psi+\sin2\psi(\cos\theta\cot\varphi-\sec\theta\tan\varphi)]}{2\sin^2\theta[\delta_j'\cos\psi(\cos\psi+\cot\varphi\sec\theta\sin\psi)-1]},
\end{equation}
\begin{equation}
    u_j^2=\frac{1}{q_j'^2}\sqrt{p_j^2\cos^2\theta+q_j^2\sin^2\theta(p_j^2\cos^2\varphi+\sin^2\varphi)},
\label{pqu2}
\end{equation}
where $\delta_j'=1-q_j'^2$ and $u_j=\sigma_j'/\sigma_j$, and $p_j=b_j/a_j$ represents the intermediate-to-long axis ratio of the $j$-th 3D Gaussian.

As mentioned previously, the misalignment between the bar major axis and the disk major axis in the observing plane is usually small and difficult to determine from the image ($\psi_{\rm bar}\approx\psi_{\rm disk}=90^\circ$). However, based on Eqs.~(\ref{pqu1})--(\ref{pqu2}), minor variations in $\psi_{\rm bar}$ can significantly affect the intrinsic shapes of the bar. Therefore, we determine $\psi_{\rm bar}$ for given values of $(\theta,\varphi)$ by adding physical constraints to the bar morphology.

First of all, the 3D luminosity density needs to remain universally positive, as Gaussians with negative luminosity are permitted during the MGE fitting process in order to fit the BP/X structure. Thus we have
\begin{equation}
    \rho(x,y,z)\ge0.
\label{rho}
\end{equation}
Then, we limited the bar axis ratio $p_{\rm bar}$ in the intrinsic $x$-$y$ plane by statistical analysis in real observations (\citealp{Marinova2007,MenendezDelmestre2007}):
\begin{equation}
    0.25 \le p_{\rm bar} \le 0.7.
\label{pbar}
\end{equation}

The BP/X-shaped structure of a barred galaxy is typically observed in the side-on view of the bar, corresponding to the $x$-$z$ plane. To quantify the prominence of the BP/X-shaped structure, we first divide the galaxy's surface brightness in the $x$-$z$ plane into 1-arcsec-width slices along the $z$-axis. For each slice at $z=z_0$, we defined the ratio of the maximum brightness (brightest $\rm1\times1\,arcsec^2$ pixel in the slice) to the central brightness ($\rm1\times1\,arcsec^2$ pixel at $x=0$) as
\begin{equation}
    \xi(z_0)=\frac{\max_x \Sigma(x_0, z_0)}{\Sigma(0, z_0)}.
\end{equation}
If the BP/X-shaped structure exists, the central pixel in the slice may not be the brightest ($\xi(z_0)>1$). The parameter $\xi$ for the entire $x$-$z$ plane is then defined as
\begin{equation}
    \xi_{xz}=\max_z \left(\frac{\max_x \Sigma(x_0, z_0)}{\Sigma(0, z_0)} \right).
\label{Xpar1}
\end{equation}
Therefore, $\xi_{xz}$ represents the maximum brightness ratio across all slices. If there is no BP/X-shaped structure, $\xi_{xz}=1$; otherwise, $\xi_{xz}>1$, with a larger value of $\xi_{xz}$ indicating a more obvious BP/X-shaped structure. Similarly, the parameters $\xi_{xy}$ and $\xi_{yz}$ in the $x$-$y$ and $y$-$z$ planes are defined as
\begin{equation}
    \xi_{xy}=\max_y \left(\frac{\max_x \Sigma(x_0, y_0)}{\Sigma(0, y_0)} \right),
\label{Xpar2}
\end{equation}
\begin{equation}
    \xi_{yz}=\max_z \left(\frac{\max_y \Sigma(y_0, z_0)}{\Sigma(0, z_0)} \right).
\label{Xpar3}
\end{equation}
Since the BP/X-shaped structure is expected to be most obvious from the side-on view of the bar, we imposed the restriction
\begin{equation}
    \xi_{xz}\ge \max(\xi_{xy},\xi_{yz}).
\label{Xshape}
\end{equation}
For the cases without BP/X-shaped structures, we have $\xi_{xz}=\xi_{xy}=\xi_{yz}=1$, which also satisfies this inequality.

Under the constraints of Eqs.~(\ref{pbar})--(\ref{Xshape}), we determine $\psi_{\rm bar}$ by minimising $|\psi_{\rm bar}-90^\circ|$ for cases without BP/X-shaped structures, and by minimising $\min(\xi_{xy},\xi_{yz})$ for cases with BP/X-shaped structures. Thus, each set of $(\theta,\varphi)$ corresponds to a unique bar position angle $\psi_{\rm bar}$ in the modelling. Fig.~\ref{mge-deprojection-different-psi} demonstrates the impact of $\psi_{\rm bar}$ on the bar morphology and how $\psi_{\rm bar}$ is determined.

After deprojecting the bar and the disk separately, the 3D luminosity density of the entire galaxy represented by MGE can be written as
\begin{equation}
    \rho_(x,y,z)=\sum_{j=1}^{N} \frac{L_j}{(\sqrt{2\pi}\sigma_j)^3 p_j q_j}\exp \left[-\frac{1}{2\sigma_j^2} \left(x^2+\frac{y^2}{p_j^2}+\frac{z^2}{q_j^2} \right)\right].
\end{equation}
The deprojected 3D luminosity distribution of Au-23-85-50, calculated under the true viewing angles $(85^\circ,50^\circ)$, is presented in Fig.~\ref{mge-deprojection}. The reconstructed luminosity distribution generally recovers the true distribution in the $x$-$y$, $x$-$z$, and $y$-$z$ planes, especially the BP/X-shaped structure.

Once the 3D luminosity distribution is determined, the mass distribution can be derived by multiplying by a constant mass-to-light ratio $M_{\star}/L$. Then the corresponding stellar potential is calculated based on the classical \citet{Chandrasekhar1969} formula (see Section 3.8 in \citealp{vdB2008} for details). For the stellar potential, the disk axis ratio $q_{\rm disk}$ (transformed from the inclination angle $\theta$), the bar azimuthal angle $\varphi$, and the mass-to-light ratio $M_{\star}/L$ are taken as free parameters in the modelling.

\subsubsection{Dark matter potential}
\label{sec3.1.2}
The density distribution of dark matter in the modelling is described by a spherical generalised NFW (gNFW) profile (\citealp{Navarro1996,Zhao1996}; see also \citealp{Barnabe2012,Cappellari2013}), which is expressed as
\begin{equation}
    \rho_{\rm DM}(r)=\frac{\rho_0}{\left(\frac{r}{R_{\rm s}}\right)^\gamma \left( 1+\frac{r}{R_{\rm s}} \right)^{3-\gamma} },
\end{equation}
where the density $\rho_0$, the scale radius $R_{\rm s}$, and the inner density slope $\gamma$ are three free parameters. $\gamma=1$ corresponds to the standard NFW profile, while $\gamma\approx0$ means a ``cored'' dark matter halo. Then, the potential of the gNFW profile can be calculated by solving Poisson's equation (see \citealp{Zhao1996}).

The above formula can be rephrased in terms of the dark matter concentration $c$, the virial mass $M_{200}$ and the inner density slope $\gamma$ by
\begin{equation}
    \rho_0=\frac{200}{3}\frac{c^3}{\zeta(c,\gamma)}\times\rho_{\rm crit},
\end{equation}
\begin{equation}
    R_{\rm s}=\left[ \frac{3}{800\pi}\frac{M_{200}}{\rho_{\rm crit}c^3} \right]^{1/3}.
\end{equation}
with
\begin{equation}
    \zeta(c,\gamma) = \int_0^c \tau^{2-\gamma} (1+\tau)^{\gamma-3} \,\mathrm{d}\tau,
\end{equation}
and the critical mass
\begin{equation}
    \rho_{\rm crit}=\frac{3H_0^2}{8\pi G},
\end{equation}
where $G$ is the gravitational constant. The dark matter concentration $c$, the virial mass $M_{200}$, and the inner density slope $\gamma$ are taken as free parameters in the modelling.

\subsubsection{Figure rotation}
\label{sec3.1.3}
When modelling a barred galaxy, it is essential to consider the figure rotation of the gravitational potential. In the rotating frame, the equations of motion and the conserved Jacobi energy $E_{\rm J}$ can be expressed as
\begin{equation}
    \ddot{\vec{r}} = -\nabla \Phi -2 \left( \vec{\Omega} \times \dot{\vec{r}} \right) -\vec{\Omega} \times \left( \vec{\Omega} \times \vec{r} \right),
\label{motion}
\end{equation}
\begin{equation}
    E_{\rm J} = \frac{1}{2} \dot{\vec{r}}^2 + \Phi - \frac{1}{2} \left| \vec{\Omega} \times \vec{r} \right|^2,
\label{Jacobi}
\end{equation}
where $\Phi$ denotes the potential, $\vec{\Omega}$ represents the angular velocity vector, and $\vec{r}$ refers to the 3D spatial vector in the rotating frame. The second and third terms of the equation correspond to the Coriolis force and the centrifugal force, respectively.

In the Cartesian coordinates $(x,y,z)$, assuming that $\Omega>0$ represents a counterclockwise rotation about the $z$-axis, Eq.~(\ref{motion}) can be converted to a scalar form:
\begin{equation}
\begin{aligned}
    \dot{x} &= v_x + \Omega y, & \quad \dot{v}_x &= -\frac{\partial \Phi}{\partial x} + \Omega v_y \\
    \dot{y} &= v_y - \Omega x, & \quad \dot{v}_y &= -\frac{\partial \Phi}{\partial y} - \Omega v_x \\
    \dot{z} &= v_z, & \quad \dot{v}_z &= -\frac{\partial \Phi}{\partial z},
\end{aligned}
\end{equation}
where $v_x$, $v_y$, $v_z$ are the velocities in the inertial frame.

Since the stellar disk and the dark matter halo are both axisymmetric in the modelling, the figure rotation of the potential will only influence the bar component. Therefore, the angular velocity $\Omega$ is equivalent to the bar pattern speed $\rm\Omega_p$ and is taken as a free parameter. Overall, there are seven free hyper-parameters in the modelling: the disk axis ratio $q_{\rm disk}$ (transformed from the inclination angle $\theta$), the bar azimuthal angle $\varphi$, the mass-to-light ratio $M_{\star}/L$, the dark matter concentration $c$, the virial mass $M_{200}$, the inner density slope of dark matter $\gamma$, and the bar pattern speed $\rm\Omega_p$.

\subsection{Orbit sampling and integration}
\label{sec3.2}
During orbit sampling and integration, the triaxial bar and the axisymmetric disk are treated as a unified triaxial system. The orbits in a triaxial system can be characterised by three integrals of motion: energy $E$, second integral $I_2$ and third integral $I_3$. Different orbit families conserve various combinations of these integrals and can be associated with distinct regions in the $x$-$z$ plane where the orbits cross perpendicularly (\citealp{Schwarzschild1993,vdB2008}). Therefore, the initial conditions of the orbits are sampled in the $x$-$z$ plane. The starting point is set as $y=0$, $v_x=v_z=0$, and $v_y=\sqrt{2[E-\Phi(x,0,z)]}$. Cartesian coordinates $(x,z)$ can be transformed into polar coordinates $(\theta,R)$, and consequently the combination $(E,\theta,R)$ can represent the initial conditions of the orbits. In this paper, we take $25\times15\times9$ combinations of $(E,\theta,R)$ to create the initial conditions. Then, by slightly perturbing the initial conditions, each orbit is dithered to $3^3$ orbits to form an orbit bundle.

As mentioned in \citet{Tahmasebzadeh2022}, retrograde orbits exhibit distinct behaviours compared to prograde orbits within a rotating triaxial potential. Therefore, it is necessary to individually sample the retrograde orbits. Then another set of initial conditions is established using the same combinations of $(E,\theta,R)$ but with the velocity changing sign $v_y=-\sqrt{2[E-\Phi(x,0,z)]}$. Finally, there are $2\times25\times15\times9=6750$ orbit bundles in total. We note that the initial velocities are sampled in the inertial frame and will be transformed into velocities in the rotating frame before integrating the orbits.

Orbital trajectories are integrated within a specified gravitational potential for $\sim200$ periods, aiming to achieve a relative accuracy of $10^{-5}$ for the conservation of the Jacobi energy $E_{\rm J}$ (see Eq.~\ref{Jacobi}). We refer to \citet{vdB2008} and \citet{Tahmasebzadeh2022} for more details of orbit sampling and integration.

\subsection{Solving orbit weights}
\label{sec3.3}
The orbit weights are solved using the non-negative least squares (NNLS; \citealp{Lawson1974}) method, taking both the stellar luminosity distribution and the kinematic data as constraints. The $\chi_{\rm err}^2$ that needs to be minimised by NNLS is expressed as
\begin{equation}
    \chi_{\rm err}^2=\chi_{\rm lum}^2+\chi_{\rm kin}^2,
\end{equation}
where $\chi_{\rm lum}^2$ and $\chi_{\rm kin}^2$ correspond to the residuals contributed by luminosity distribution and kinematic data, respectively. For the luminosity distribution, we use both the 2D surface brightness fitted by the MGE method and the 3D luminosity density deprojected from the 2D MGE as constraints. The 2D surface brightness is stored in the same apertures as the kinematic data, with $S_i$ representing the surface brightness in the $i$-th aperture. The 3D space is divided into 360 cells, with $\gamma_m$ representing the luminosity of the $m$-th cell. We set the relative errors of $S_i$ to be $1\%$ and $\gamma_m$ to be $2\%$. Thus we have
\begin{equation}
    \chi_{\rm lum}^2=\sum_{i=1}^{N_{\rm obs}}\left(\frac{S_i^*-S_i}{0.01S_i}\right)^2+\sum_{\rm m=1}^{M}\left(\frac{\gamma_m^*-\gamma_m}{0.02\gamma_m}\right)^2,
\end{equation}
where $N_{\rm obs}$ is the number of 2D apertures and $M$ is the number of 3D cells. For the kinematic data, we fit the Gauss–Hermite coefficients \citep{Gerhard1993,vdMarel1993,Rix1997} of the line-of-sight velocity distribution. IFS data usually provide maps of velocity $V$, velocity dispersion $\sigma$, and higher-order coefficients $h_3$ and $h_4$. The first and second orders of Gauss-Hermite coefficients and their errors $h_1$, $h_2$, $\Delta h_1$, and $\Delta h_2$ can be converted from $V$, $\sigma$, $\Delta V$ and $\Delta \sigma$. Thus we have
\begin{equation}
\begin{split}
    \chi_{\rm kin}^2=\sum_{i=1}^{N_{\rm obs}}\left[ \left(\frac{h^*_{1i}-h_{1i}}{\Delta h_{1i}}\right)^2+\left(\frac{h^*_{2i}-h_{2i}}{\Delta h_{2i}} \right)^2+\right.&\\
    \quad\quad\quad\quad\ \
    \left.\left(\frac{h^*_{3i}-h_{3i}}{\Delta h_{3i}} \right)^2+\left(\frac{h^*_{4i}-h_{4i}}{\Delta h_{4i}} \right)^2\right].
\label{chi2nnls}
\end{split}
\end{equation}
The variables marked with a `$*$' indicate the model fittings, those with a `$\Delta$' represent the errors of the input data and those without a marker denote the input data. With the derived orbit libraries and their corresponding weights, a barred orbit-superposition model is created for a given set of free parameters.

\section{Searching for the best-fitting model and comparing with the truth}
\label{sec4}
\subsection{The best-fitting model and model uncertainties}
\label{sec4.1}
\begin{figure*}
    \centering
    \includegraphics[width=17.8cm]{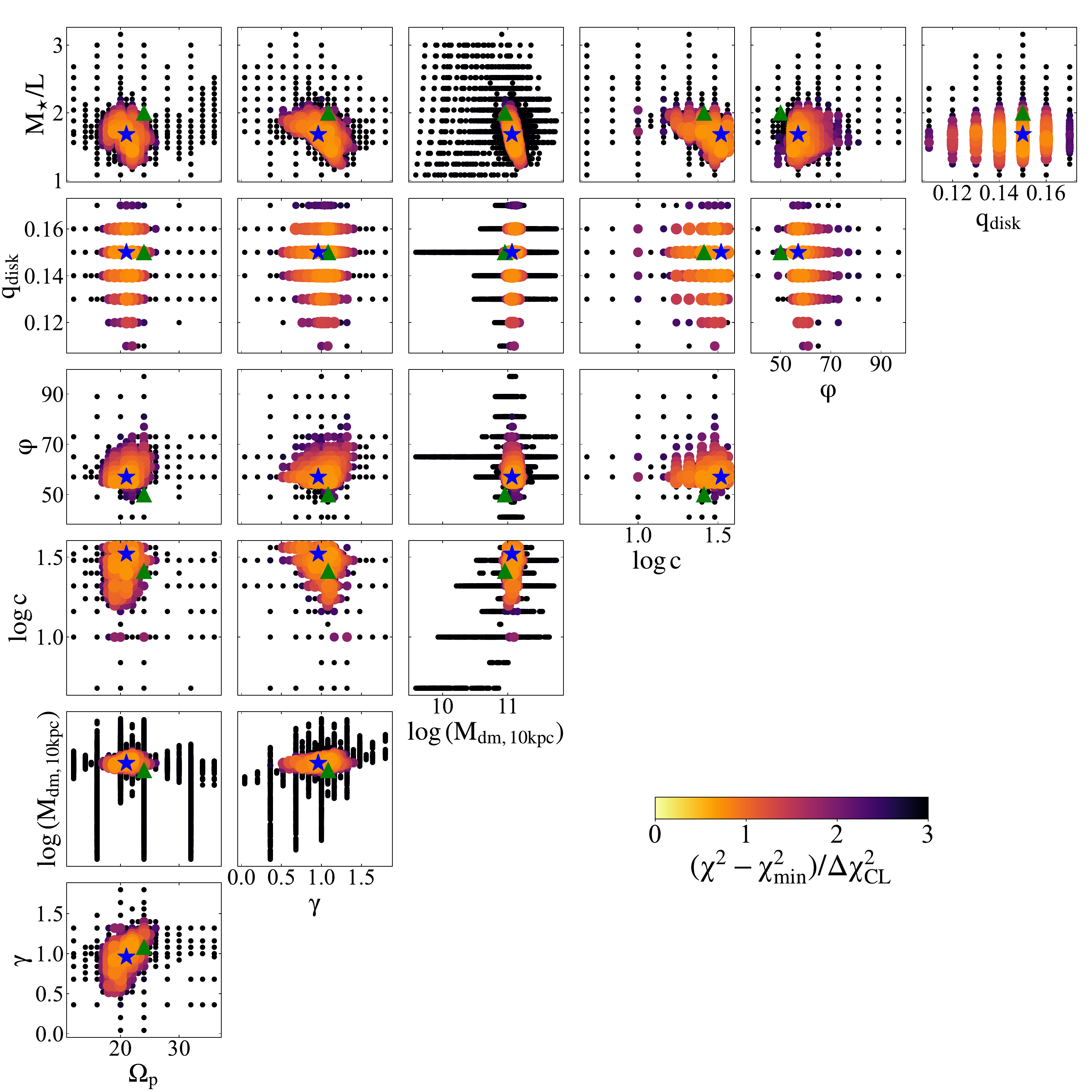}
    \caption{Parameter space we explored for Au-23-85-50. The seven free parameters are the disk axis ratio $q_{\rm disk}$ (transformed from the inclination angle $\theta$), the bar azimuthal angle $\varphi$, the mass-to-light ratio $M_{\star}/L$, the dark matter concentration $c$, the dark matter mass $M_{\rm dm,10kpc}$ within $\rm10\,kpc$ (transformed from the virial mass $M_{\rm 200}$), the inner density slope of dark matter $\gamma$, and the bar pattern speed $\rm\Omega_p$. The largest blue pentacle represents the best-fitting model with minimum chi-square $\chi^2_{\rm min}$, and other coloured dots represent models within the $3\sigma$ confidence level, with their $\chi^2$ values indicated by the colour bar. The small black dots denote models outside of the $3\sigma$ confidence level. The green triangle indicates the true values of the free parameters, with the dark matter parameters $c$ and $\gamma$ are obtained by fitting the gNFW profile using dark matter particles located within $\rm 10\,kpc$.}
    \label{parameter-grids}
\end{figure*}
\begin{figure*}
    \centering
    \includegraphics[width=16cm]{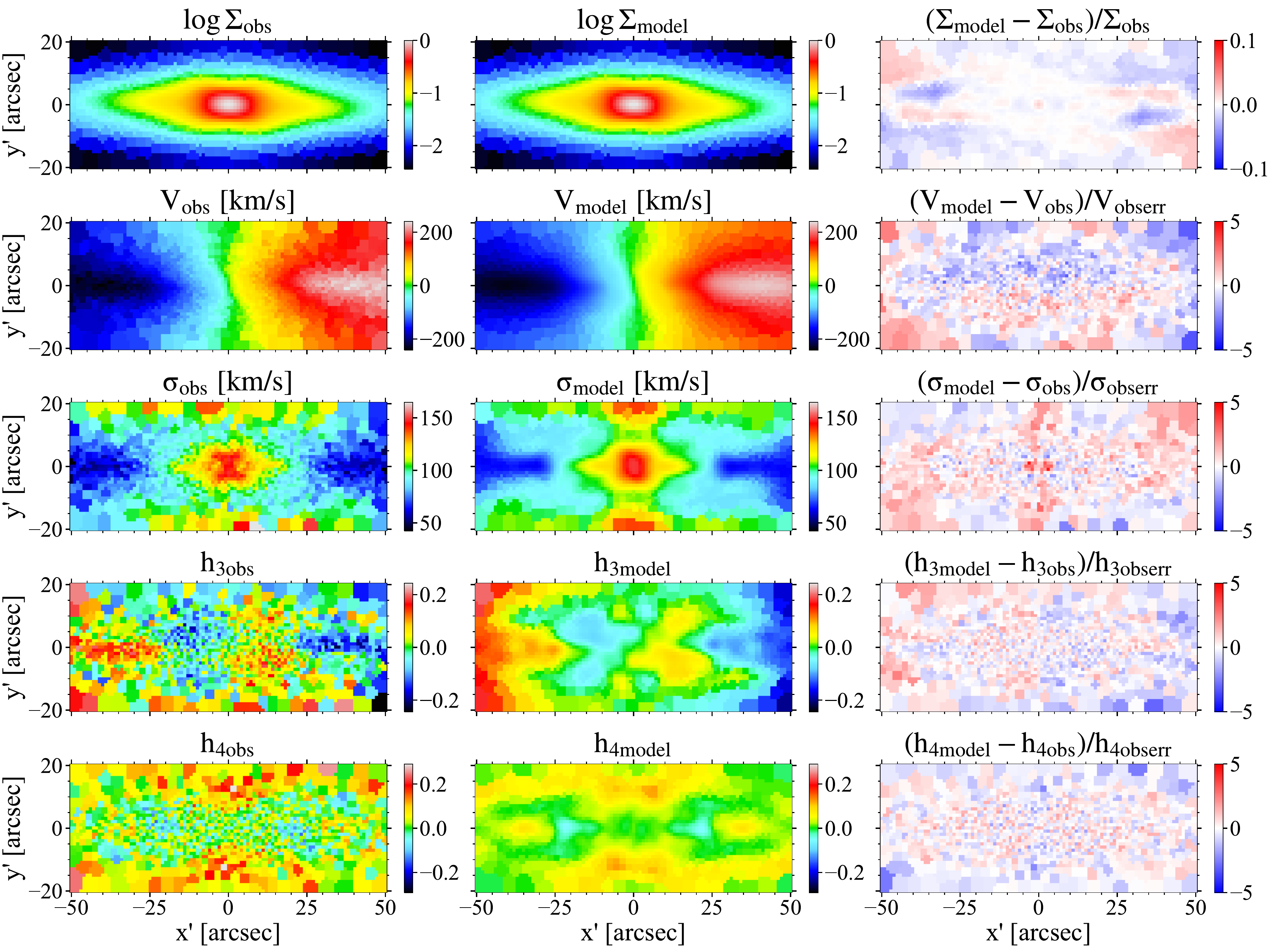}
    \caption{Mock surface brightness, mock kinematic maps, and best-fitting model for Au-23-85-50. From top to bottom: Logarithmic normalised surface brightness $\log\Sigma$, mean velocity $V$, velocity dispersion $\sigma$, third-order Gauss-Hermite coefficient $h_3$, and fourth-order Gauss-Hermite coefficient $h_4$. From left to right: Mock observations, model fittings, and residuals. The residuals represent the relative deviations of the surface brightness and the standardised residuals of the kinematics.}
    \label{best-fitting-kinematics}
\end{figure*}

As mentioned in Sect.~\ref{sec3.1}, we have seven free hyper-parameters in the modelling: $q_{\rm disk}$ (transformed from $\theta$), $\varphi$, $M_{\star}/L$, $c$, $M_{200}$, $\gamma$, and $\rm\Omega_p$. For each set of parameters, we solve orbit weights to determine the model by minimising $\chi^2_{\rm err}=\chi^2_{\rm lum}+\chi^2_{\rm kin}$. Here, we evaluate the goodness of each model by calculating the difference between model-fitted and mock kinematic maps directly, which can be written as
\begin{equation}
\begin{split}
    \chi^2=\sum_{i=1}^{N_{\rm obs}}\left[ \left(\frac{V^*_i-V_i}{\Delta V_i}\right)^2+\left(\frac{\sigma^*_i-\sigma_i}{\Delta \sigma_i} \right)^2+\right.&\\
    \left.\left(\frac{h^*_{3i}-h_{3i}}{\Delta h_{3i}} \right)^2+\left(\frac{h^*_{4i}-h_{4i}}{\Delta h_{4i}} \right)^2\right],\\
\end{split}
\label{chi2kin}
\end{equation}
where the markers of the variables have the same meaning as Eq.~(\ref{chi2nnls}). In principle, $\chi^2$ should be highly correlated with $\chi^2_{\rm kin}$. However, if the line-of-sight velocity distribution deviates significantly from a Gaussian distribution, $\chi^2_{\rm kin}$ and $\chi^2$ may not be strongly correlated. Since $\chi^2$ directly reflects how the model matches the data, it can better represent the goodness of the model. In practice, luminosity distributions are much easier to fit than kinematic data, so $\chi_{\rm lum}^2$ is a small value and is not taken into account in the goodness evaluation.

We take an iterative process to search for the best-fitting model in the parameter space. We start by making initial guesses about the free parameters and construct initial models. Models with a relatively lower $\chi^2$ are selected, and new models are created by walking a few steps in the parameter space around the selected models. This process is repeated until we obtain a minimum $\chi^2$, with all surrounding models calculated in the parameter space. Finally, we find the model with the minimum chi-square $\chi^2_{\rm min}$, which is defined as the best-fitting model. We take the best-fitting model as the default model in our analysis.

The confidence level of the modelling, which determines the uncertainties of the free parameters, is calculated by perturbing the kinematic maps and resolving the orbit weights. For each mock data set, the kinematic maps $(V_i,\sigma_i,h_{3i},h_{4i})$ are perturbed 1000 times with their error maps $(\Delta V_i,\Delta\sigma_i,\Delta h_{3i},\Delta h_{4i})$: $V_i'=V_i+N(0,\Delta V_i)$, $\sigma_i'=\sigma_i+N(0,\Delta \sigma_i)$, $h_{3i}'=h_{3i}+N(0,\Delta h_{3i})$, and $h_{4i}'=h_{4i}+N(0,\Delta h_{4i})$, where $N(0,\Delta x)$ represents the Gaussian distribution with centre 0 and dispersion $\Delta x$. Then the perturbed kinematic maps are point-symmetrised in the same way as the original maps. Fixing the gravitational potential determined by the best-fitting parameters, we resolve the orbit weights under the constraints of the perturbed kinematic maps and obtain 1000 new $\chi^2$. The distribution of these $\chi^2$ is Gaussian-like. Therefore, we interpret the fluctuation of these 1000 $\chi^2$ values within the $\pm1\sigma$, $\pm2\sigma$, and $\pm3\sigma$ regions of this Gaussian-like distribution as the $1\sigma$ ($68\%$, $\Delta\chi^2_{\rm CL}$), $2\sigma$ ($95\%$, $2\Delta\chi^2_{\rm CL}$), and $3\sigma$ ($>99\%$, $3\Delta\chi^2_{\rm CL}$) confidence levels of the modelling, respectively. The uncertainties of the free parameters are defined as the range of model values within the $1\sigma$ confidence level ($\chi^2-\chi^2_{\rm min}\le\Delta\chi^2_{\rm CL}$).

\subsection{Comparing the model with the truth}
\label{sec4.2}
In order to evaluate how our models match the truth, we calculate the corresponding true values of the free parameters from the simulations. Since our modelling can only constrain the galaxy properties within the data coverage ($\le\rm10\,kpc$), we convert the virial mass $M_{\rm 200}$ into a more representative parameter $M_{\rm dm,10kpc}$. Thus, the parameters that we compare with the truth become $q_{\rm disk}$, $\varphi$, $M_{\star}/L$, $c$, $M_{\rm dm,10kpc}$, $\gamma$, and $\rm\Omega_p$.

The true disk axis ratio $q_{\rm disk,T}$ is calculated by projecting the mock galaxy onto the $x$-$z$ plane, separating the disk component through GALFIT and MGE fitting, and measuring the axis ratio of the disk isophote at $x=\rm60\,arcsec$ ($\rm 1\,arcsec=0.2\,kpc$), which follows the same way as used for the models (see Sect.~\ref{sec3.1.1}). The true values of the bar azimuthal angle $\varphi_{\rm T}$ and the mass-to-light ratio $(M_{\star}/L)_{\rm T}$ are predefined when creating the mock data sets. The true dark matter mass, $M_{\rm dm,10kpc,T}$, is obtained directly by summing all dark matter particle masses within $\rm10\,kpc$. The true dark matter concentration $c_{\rm T}$ and the true inner density slope $\gamma_{\rm T}$ are derived by fitting the true dark matter density profile within $\rm10\,kpc$ to the gNFW profile.

The true values of the bar pattern speeds $\rm\Omega_T$ come from measurements in different works. For Au-23, we adopt the TW method measurement $\rm\Omega_T=24\,km\,s^{-1}\,kpc^{-1}$ as the true value ($5\%$ accuracy; \citealp{Fragkoudi2020}). For Au-28, the true pattern speed is calculated as an annular average within the bar radius $\Omega_{\rm T}=\Delta\langle\theta\rangle/\Delta t$ using high-cadence snapshots in simulations \citep{Fragkoudi2021}. This results in a value of $\rm\Omega_T=40\,km\,s^{-1}\,kpc^{-1}$ with a systematic uncertainty of $\sim\rm3\,km\,s^{-1}\,kpc^{-1}$, which is caused by oscillations in the pattern speeds at different snapshots. For Au-18, the measurements from the TW method \citep{Fragkoudi2020} and the high-cadence snapshots \citep{Fragkoudi2021} provide a consistent pattern speed $\rm\Omega_T=28\,km\,s^{-1}\,kpc^{-1}$, with an uncertainty of $\sim\rm1\,km\,s^{-1}\,kpc^{-1}$.

The parameter space we explored for Au-23-85-50 and the true values of the free parameters are shown in Fig.~\ref{parameter-grids}. The disk axis ratio $q_{\rm disk}$ agrees with the truth, while the bar azimuthal angle $\varphi$ is slightly overestimated by $7^\circ$. The mass-to-light ratio recovered by the model $M_{\star}/L=1.68^{+0.28}_{-0.40}$ is $16\%$ lower than the true value $(M_{\star}/L)_{\rm T}=2$, while the dark matter halo in the model is more concentrated than the truth, and its mass $M_{\rm dm,10kpc}$ is overestimated by $28\%$. This might be attributed to the deviation of real dark matter halo from spherical gNFW profile. The mass profiles for Au-23-85-50 are presented in Fig.~\ref{mass-profiles}. Despite the large degeneracy between dark matter and stellar mass, the total mass within the bar region ($\rm\lesssim5\,kpc$), which directly affects the rotation of the bar, is well recovered. The model-recovered bar pattern speed $\rm\Omega_p=21^{+4}_{-4}\,km\,s^{-1}\,kpc^{-1}$ matches the true value $\rm\Omega_T=24\,km\,s^{-1}\,kpc^{-1}$.

In Fig.~\ref{best-fitting-kinematics}, we show the mock surface brightness, mock kinematic maps, and best-fitting model for Au-23-85-50. The model reproduces the features of the kinematics, especially the warp in the central region of the velocity map, which is caused by the rotation of the bar. Similar to Fig.~\ref{best-fitting-kinematics}, Fig.~\ref{best-fitting-kinematics-end-on} presents the recoveries of surface brightness and kinematic maps for a bar end-on case, Au-23 with $(\theta_{\rm T},\varphi_{\rm T})=(89^\circ,10^\circ)$.

\section{Model-recovered bar pattern speeds}
\label{sec5}
In this section, we present the model-recovered bar pattern speeds $\rm\Omega_p$ for all mock data sets and compare them with the true pattern speeds $\rm\Omega_T$. As described in Sect.~\ref{sec2.2}, our mock data sets include 12 bar side-on cases ($\varphi_{\rm T}\ge50^\circ$) and four bar end-on cases ($\varphi_{\rm T}\le30^\circ$). Their corresponding results are presented in Sects.~\ref{sec5.1} and~\ref{sec5.2}, respectively.

\subsection{Model-recovered pattern speeds for side-on bars}
\label{sec5.1}
\begin{figure*}
    \centering
    \includegraphics[width=17.8cm]{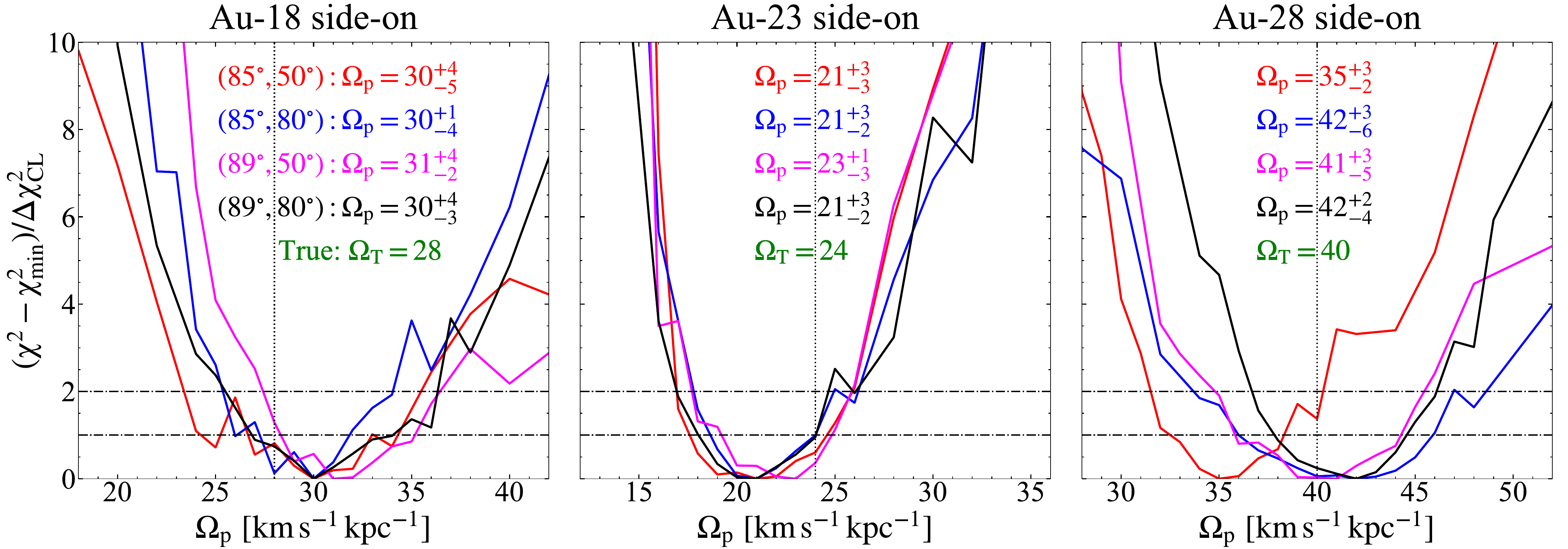}
    \caption{Marginalised chi-squares $\chi^2_{\rm norm}=(\chi^2-\chi^2_{\rm min})/\Delta\chi^2_{\rm CL}$ as a function of model-recovered bar pattern speeds $\rm\Omega_p$ for bar side-on cases ($\varphi_{\rm T}\ge50^\circ$). From left to right, the panels show the results for Au-18, Au-23, and Au-28. Each fold line represents a complete set of modelling runs. The red, blue, magenta, black fold lines denote the viewing angles of the mock data sets $(\theta_{\rm T},\varphi_{\rm T})=(85^\circ,50^\circ)$, $(85^\circ,80^\circ)$, $(89^\circ,50^\circ)$, and $(89^\circ,80^\circ)$, respectively. The horizontal dashed lines represent $1\sigma$ ($68\%$, $\chi^2_{\rm norm}=1$) and $2\sigma$ ($95\%$, $\chi^2_{\rm norm}=2$) confidence levels. The vertical dotted lines denote the true bar pattern speeds, with their values shown by the green annotations.}
    \label{chi2-vs-omega-side-on-cases}
\end{figure*}

Figure ~\ref{chi2-vs-omega-side-on-cases} illustrates the influence of model-recovered bar pattern speeds $\rm\Omega_p$ on marginalised chi-squares $\chi^2_{\rm norm}=(\chi^2-\chi^2_{\rm min})/\Delta\chi^2_{\rm CL}$ for the 12 bar side-on cases. At given $\rm\Omega_p$, we use the model with minimum $\chi^2_{\rm norm}$ in the parameter space to plot the figure. In all cases, the bar pattern speeds are well recovered, with the distributions of normalised chi-squares $\chi^2_{\rm norm}$ centring near the true pattern speeds. For the best-fitting values of $\rm\Omega_p$, five cases slightly underestimate the pattern speeds by less than $\rm 5\,km\,s^{-1}\,kpc^{-1}$, while the other seven cases slightly overestimate the pattern speeds by less than $\rm 3\,km\,s^{-1}\,kpc^{-1}$. This results in a percentage accuracy of $13\%$ ($|\rm\Delta\Omega_p/\Omega_T|\le13\%$), and is comparable with the pattern speed estimations for non-edge-on galaxies derived from the orbit-superposition method (\citealp{Tahmasebzadeh2022}; $|\rm\Delta\Omega_p/\Omega_T|\lesssim10\%$) and the TW method (\citealp{Zou2019}; $|\rm\Delta\Omega_p/\Omega_T|\lesssim10\%$). For 10 of 12 cases, the true pattern speeds $\rm\Omega_T$ lie within the $1\sigma$ ($68\%$) confidence levels, while for the remaining two cases, $\rm\Omega_T$ falls between the $1\sigma$ and $2\sigma$ ($95\%$) confidence levels. This indicates that our model results are generally consistent with the probabilistic estimates of confidence levels. For all 12 cases, the average model uncertainty is equal to $10\%$.

\subsection{Model-recovered pattern speeds for end-on bars}
\label{sec5.2}
\begin{figure*}
    \centering
    \includegraphics[width=17.8cm]{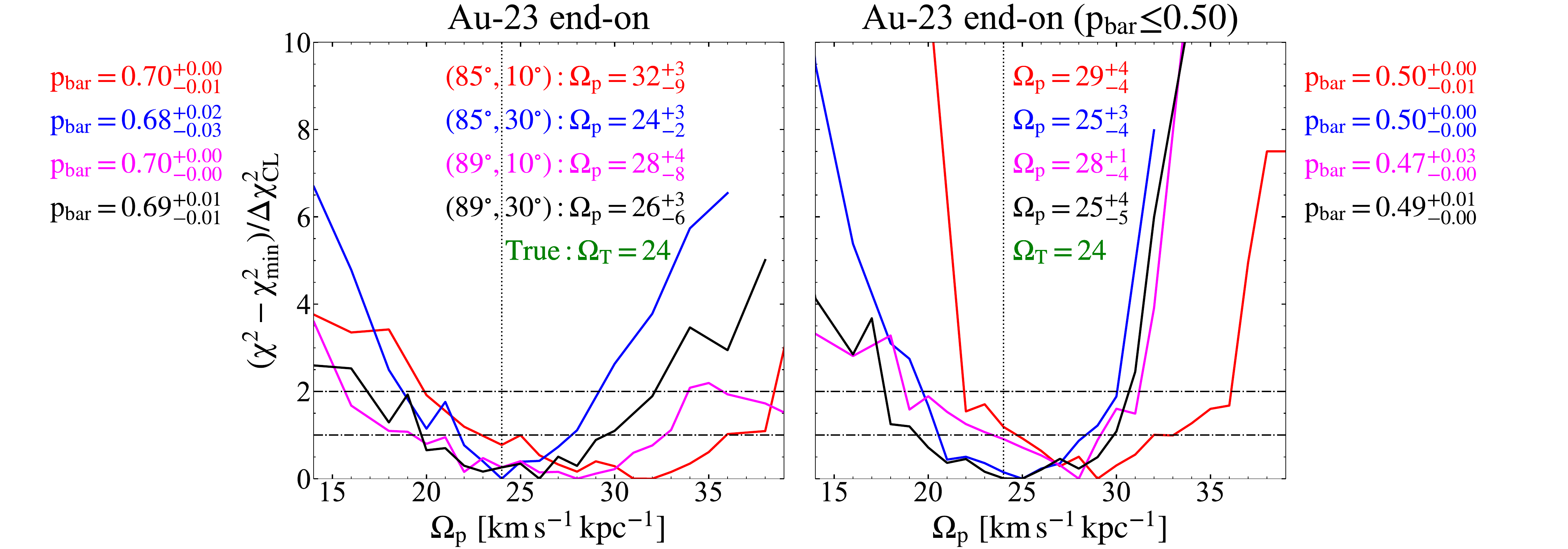}
    \caption{Marginalised chi-squares $\chi^2_{\rm norm}=(\chi^2-\chi^2_{\rm min})/\Delta\chi^2_{\rm CL}$ as a function of model-recovered bar pattern speeds $\rm\Omega_p$ for bar end-on cases ($\varphi_{\rm T}\le30^\circ$). Left panel: the results from the original models. Right panel: the results from the revised models with stronger assumptions that the bars are strong ($p_{\rm bar}\le0.50$). Each fold line represents a complete set of modelling runs. The red, blue, magenta, black fold lines denote the viewing angles of the mock data sets $(\theta_{\rm T},\varphi_{\rm T})=(85^\circ,10^\circ)$, $(85^\circ,30^\circ)$, $(89^\circ,10^\circ)$, and $(89^\circ,30^\circ)$, respectively. The horizontal dashed lines represent $1\sigma$ ($68\%$, $\chi^2_{\rm norm}=1$) and $2\sigma$ ($95\%$, $\chi^2_{\rm norm}=2$) confidence levels, and the vertical dotted lines denote the true bar pattern speed of Au-23, with its value shown by the green annotations. The coloured annotations outside each panel represent the values of bar axis ratio $p_{\rm bar}$ in the models.}
    \label{chi2-vs-omega-end-on-cases}
\end{figure*}

Similar to Figure~\ref{chi2-vs-omega-side-on-cases}, the left panel of Fig.~\ref{chi2-vs-omega-end-on-cases} illustrates the influence of model-recovered bar pattern speeds $\rm\Omega_p$ on marginalised chi-squares $\chi^2_{\rm norm}=(\chi^2-\chi^2_{\rm min})/\Delta\chi^2_{\rm CL}$ for Au-23 with end-on bars. The model-recovered bar pattern speeds are $\rm\Omega_p=32^{+3}_{-9}$, $24^{+3}_{-2}$, $28^{+4}_{-8}$, and $26^{+3}_{-6}\,\rm km\,s^{-1}\,kpc^{-1}$ for viewing angles $(\theta_{\rm T},\varphi_{\rm T})=(85^\circ,10^\circ)$, $(85^\circ,30^\circ)$, $(89^\circ,10^\circ)$, and $(89^\circ,30^\circ)$, respectively. For the case with $(\theta_{\rm T},\varphi_{\rm T})=(85^\circ,30^\circ)$, the uncertainties of $\rm\Omega_p$ are comparable with the side-on cases of Au-23, while for the case with $(\theta_{\rm T},\varphi_{\rm T})=(89^\circ,30^\circ)$, $\rm\Omega_p$ exhibits slightly larger uncertainties. For cases with $\varphi_{\rm T}=10^\circ$, where the bar is nearly perpendicular to the observing plane, the recoveries of $\rm\Omega_p$ are the poorest, with uncertainties approaching $30\%$. In these end-on cases, the best-fitting bar azimuthal angles $\varphi$ range from $50^\circ$ to $60^\circ$, significantly deviating from the true values $\varphi_{\rm T}=10^\circ$ and $30^\circ$, and the corresponding bar axis ratios $p_{\rm bar}$ approach the maximum value of 0.70 defined in Eq.~(\ref{pbar}) across all cases. This implies that the models tend to interpret end-on strong bars as side-on weak bars.

In order to evaluate how model assumptions affect $\rm\Omega_p$, we further restrict the bars to be strong bars ($p_{\rm bar}\le0.50$). We keep all other constraints unchanged (see Sect.~\ref{sec3.1.1}) and reconstruct the models using the same mock data sets. The results of these revised models are shown in the right panel of Fig.~\ref{chi2-vs-omega-end-on-cases}. The recovered bar pattern speeds of these models are $\rm\Omega_p=29^{+4}_{-4}$, $25^{+3}_{-4}$, $28^{+1}_{-4}$, and $25^{+4}_{-5}$ $\rm km\,s^{-1}\,kpc^{-1}$, respectively. The $\chi^2_{\rm norm}$ distributions of the revised models become more concentrated compared to those of the original models, with the uncertainties of $\rm\Omega_p$ slightly reduced in the cases with $(\theta_{\rm T},\varphi_{\rm T})=(85^\circ,30^\circ)$ and $(89^\circ,10^\circ)$, and significantly reduced in the case with $(\theta_{\rm T},\varphi_{\rm T})=(89^\circ,10^\circ)$. These results lead to an average uncertainty of $14\%$, which is slightly larger than that of the side-on cases. The true pattern speed $\rm\Omega_T$ falls within the $1\sigma$ ($68\%$) confidence levels for three of four cases and within the $2\sigma$ ($95\%$) confidence levels for all four cases.

\begin{figure*}
    \centering
    \includegraphics[width=17.8cm]{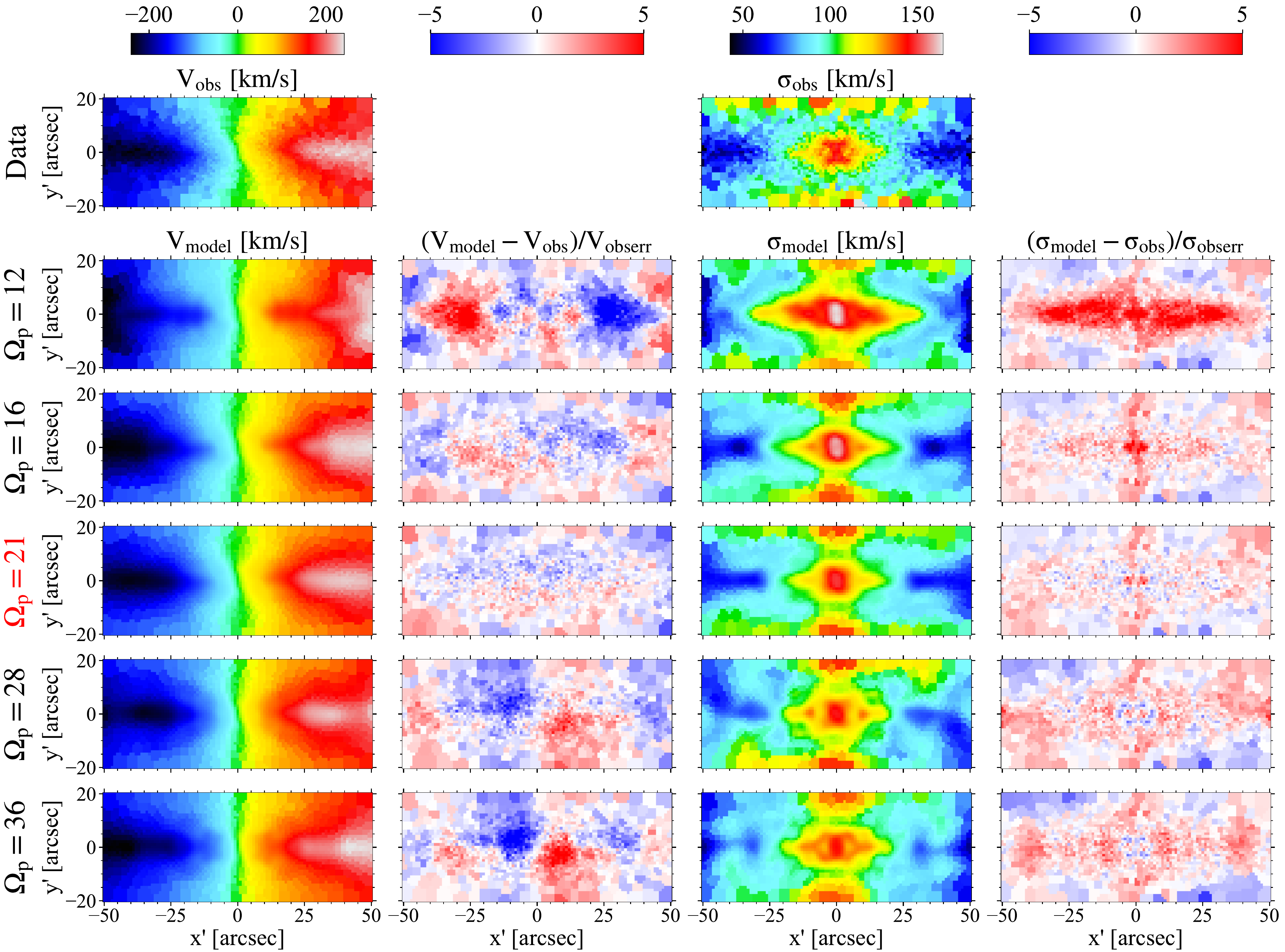}
    \caption{Influence of $\rm\Omega_p$ on the mean velocity and velocity dispersion maps for Au-23-85-50. The topmost two panels represent the mock mean velocity $V_{\rm obs}$ and velocity dispersion $\sigma_{\rm obs}$ maps. From the second row to the sixth row, each row denotes a model. The fourth row with $\rm\Omega_p=21\,km\,s^{-1}\,kpc^{-1}$ represents the best-fitting model, while the remaining rows represent the model fitting results with only a change in $\rm\Omega_p$. From left to right: mean velocity $V_{\rm model}$, standardised residual of mean velocity $(V_{\rm model}-V_{\rm obs})/V_{\rm obs}$, velocity dispersion $\sigma_{\rm model}$, and standardised residual of velocity dispersion $(\sigma_{\rm model}-\sigma_{\rm obs})/\sigma_{\rm obs}$.}
    \label{kinematics-different-omega}
\end{figure*}

\subsection{Model constraints on bar pattern speeds}
\label{sec5.3}
The test results in Sects.~\ref{sec5.1} and~\ref{sec5.2} indicate that the recovery of the bar pattern speeds $\rm\Omega_p$ depends significantly on the viewing angles $(\theta_{\rm T},\varphi_{\rm T})$ of the mock data sets. Side-on views ($\varphi_{\rm T}\ge50^\circ$) provide enough kinematic information about bar rotations, and thus we can better constrain $\rm\Omega_p$. Furthermore, for Au-18 and Au-23, the obvious BP/X-shaped structures in the mock surface brightness can effectively constrain the intrinsic bar morphology using Eqs.~(\ref{pbar})--(\ref{Xshape}), which may also contribute to the recovery of $\rm\Omega_p$. In Fig.~\ref{kinematics-different-omega}, we show the influence of $\rm\Omega_p$ on the kinematic maps for Au-23-85-50, with all other parameters unchanged. With increasing or decreasing $\rm\Omega_p$ away from the best-fitting value, the $\chi^2$ in the mean velocity and the velocity dispersion maps along the bar obviously increase. This explains why $\rm\Omega_p$ of the bar side-on cases can be well constrained.

However, for the bar end-on cases ($\varphi_{\rm T}\le30^\circ$), we have less information about the bar rotations. Since the modelling constrains $\rm\Omega_p$ with projected kinematics, cases with bars closer to being perpendicular to the observing plane ($\theta_{\rm T}\approx90^\circ$ and $\varphi_{\rm T}\approx0^\circ$) lead to weaker constraints of $\rm\Omega_p$. This explains why the mock data set with $(\theta_{\rm T},\varphi_{\rm T})=(89^\circ,10^\circ)$ results in the poorest recovery of the end-on cases, whereas the mock data set with $(\theta_{\rm T},\varphi_{\rm T})=(85^\circ,30^\circ)$ achieves the best recovery.

As described in Sect.~\ref{sec5.2}, without further restrictions on the bar morphology, our models might treat end-on strong bars as side-on weak bars, which means the bar azimuthal angles $\varphi$ are poorly recovered. Nevertheless, except for the most extreme case with $(\theta_{\rm T},\varphi_{\rm T})=(89^\circ,10^\circ)$, the recovery of $\rm\Omega_p$ is not significantly affected by the poor recovery of $\varphi$. In the modelling, $\rm\Omega_p$ is equal to the projected velocity at the end of the bar ($\approx V_{\rm bar}\sin\varphi$) divided by the projected bar length ($\approx L_{\rm bar}\sin\varphi)$, and thus is weakly correlated with $\varphi$. We present a figure similar to Fig.~\ref{kinematics-different-omega} but for the original model with $(\theta_{\rm T},\varphi_{\rm T})=(89^\circ,10^\circ)$ in Fig.~\ref{kinematics-different-omega-end-on}. The deviations of $\rm\Omega_p$ from the best-fitting value do not significantly enlarge $\chi^2$ in the mean velocity and the velocity dispersion maps.

\section{Discussion}
\label{sec6}
\subsection{The influence of separating radii $R_0$ on bar pattern speeds}
\label{sec6.1}
As mentioned in Sect.~\ref{sec3.1.1}, we determine a separating radius $R_0$ for each mock observation to distinguish between the disk Gaussians and the bar Gaussians in the MGE fitting. These separating radii are derived from the photometric decomposition algorithm GALFIT, which might exhibit some uncertainties and thus influence our results. We test the impact of $R_0$ on the bar pattern speed $\rm\Omega_p$ for three cases: Au-18-89-50 (side-on bar), Au-28-85-80 (side-on bar), and Au-23-85-30 (end-on bar with $p_{\rm bar}\le0.50$). For each case, we adopt a smaller ($R_0=0.6R_{0,\rm default}$) and a larger ($R_0=1.3R_{0,\rm default}$) separating radius, and reconstruct the models with all other constraints unchanged. We compare the results using smaller and larger radii with the default results in Fig.~\ref{chi2-vs-omega-test-R0}. For the two bar side-on cases, decreasing $R_0$ significantly increases the uncertainties of $\rm\Omega_p$. This occurs because smaller $R_0$ values result in shorter bars, and thus weaken the constraints on $\rm\Omega_p$ from kinematic data. Conversely, larger $R_0$ values have minimal impacts on the recoveries of $\rm\Omega_p$. This is because line-of-sight velocities at the bar ends are comparable to those in the disks; thus longer bars with similar $\rm\Omega_p$ can still reproduce the observed kinematics. For the bar end-on case under the strong bar assumption, variations in $R_0$ (either decreasing or increasing) do not result in worse recoveries. This indicates that the choice of $R_0$ has greater impact in side-on cases than in end-on cases. Nevertheless, for all test results, the $2\sigma$ ($95\%$) confidence levels of $\rm\Omega_p$ consistently encompass the true pattern speeds $\rm\Omega_T$.

\subsection{The influence of BP/X-shaped structures and dark matter halos}
\label{sec6.2}
In this work, we model two galaxies with obvious BP/X-shaped structures (Au-18, Au-23) and one without such features (Au-28). As shown in Fig.~\ref{chi2-vs-omega-side-on-cases}, Au-28 achieves a comparable precision in recovering $\rm\Omega_p$ compared to Au-18 and Au-23. To better understand the influence of BP/X-shaped structures on our model results, we reconstruct two models for Au-23-85-50 and Au-23-85-80, ignoring BP/X-shaped structures (do not use negative Gaussians; see Sect.~\ref{sec3.1.1}) and keeping other constraints unchanged. The comparison between models with and without considering BP/X-shaped structures is shown in Fig.~\ref{chi2-vs-omega-test-peanut-DMhalo}. For both cases, ignoring BP/X-shaped structures does not degrade the recoveries of $\rm\Omega_p$. The uncertainties of the bar azimuthal angles $\varphi$ increase significantly for Au-23-85-50 but remain consistent for Au-23-85-80. This difference occurs because the deprojection of BP/X-shaped structures is restricted to a small region when $\varphi\lesssim50^\circ$.

As indicated in Figs.~\ref{parameter-grids} and~\ref{mass-profiles}, large degeneracies exist between stellar mass and dark matter mass in our models. To assess the influence of these degeneracies, we reconstruct two models with constrained dark matter halos for Au-23-85-50 and Au-23-85-80, where the differences between model-recovered and true dark matter masses within $\rm10\,kpc$ are limited to be less than $10\%$ ($0.9M_{\rm dm,10kpc}^{\rm true}<M_{\rm dm,10kpc}^{\rm model}<1.1M_{\rm dm,10kpc}^{\rm true}$), with all other constraints unchanged. For both cases, the recoveries of $\rm\Omega_p$ and $\varphi$ do not change significantly. This supports that bar kinematics are reliably recovered when the total mass within bar regions is well constrained, and inaccurate dark matter halos do not significantly impact our results.

\subsection{Application to real observations in the future}
\label{sec6.3}
Before applying the method to edge-on galaxies in real observations, we first need to judge if a bar is present when the host galaxy does not exhibit an obvious BP/X-shaped structure in its image. Since our models might interpret end-on strong bars as side-on weak bars (as described in Sect.~\ref{sec5.2}), more accurate estimates of bar azimuthal angles can help us resolve this degeneracy.

It is possible to identify a bar from both the stellar kinematics and gas distribution perspectives. Line-of-sight velocity distributions with a high-velocity tail (corresponding to positive $h_3$--$V$ correlation) are bar diagnostics (e.g. \citealp{Bureau2005,Molaeinezhad2016,LiZhaoyu2018}), while the line-of-sight velocity distributions of the disks have low-velocity tails (corresponding to negative $h_3$--$V$ correlation). For example, Au-23 has such features from both side-on views (see Figs.~\ref{kinematic-data} and~\ref{best-fitting-kinematics}) and end-on views (see Fig.~\ref{best-fitting-kinematics-end-on}). Furthermore, previous studies have shown that the molecular gas traced by CO and the ionised gas traced by H$\rm\alpha$ peak not only in the galaxy's central region but also at the end of the bar (e.g. \citealp{Regan1999,Verley2007,DiazGarcia2020,FraserMcKelvie2020,Maeda2020}). These features can also be taken as bar diagnostics.  

The velocity dispersion $\sigma$ in the observed maps can help us distinguish end-on strong bars from side-on weak bars. When the bar is end-on, $\sigma$ increases significantly in the central regions due to the corotating orbits elongated along the bar (e.g. \citealp{Iannuzzi2015}; see also \citealp{Athanassoula1992,Bureau1999}). These diagnostics were used to identify two edge-on galaxies that host strong end-on bars, IC 1711 and IC 5244, in the GECKOS survey \citep{FraserMcKelvie2024}.

In real observations, the stellar mass-to-light ratio $M_{\star}/L$ might exhibit obvious gradients, deviating from the constant $M_{\star}/L$ assumed in our modelling. This can increase the degeneracies between stellar mass and dark matter mass. Although these degeneracies may not significantly affect estimates of bar pattern speeds $\rm\Omega_p$ (as discussed in Sect.~\ref{sec6.2}), we can address this through two approaches. First, we can incorporate the $M_{\star}/L$ gradients derived from stellar population synthesis models into our dynamical models. Second, if available, we can utilise $\rm3.6\,\mu m$ galaxy images from the Spitzer Space Telescope Infrared Array Camera \citep{Werner2004,Fazio2004} to trace stellar mass distributions, as near-infrared wavelengths are less affected by dust extinction and correlate more directly with older stellar populations than optical wavelengths.

\section{Summary}
\label{sec7}
We have developed an orbit-superposition method suitable for edge-on barred galaxies based on \citet{Tahmasebzadeh2021,Tahmasebzadeh2022} that is able to estimate the bar pattern speeds $\rm\Omega_p$. We selected three simulated galaxies, Au-18, Au-23, and Au-28, from the Auriga simulations and created MUSE-like mock data sets with different viewing angles $(\theta_{\rm T},\varphi_{\rm T})$, including 12 mock data sets with side-on bars ($\varphi_{\rm T}\ge50^\circ$) and 4 mock data sets with end-on bars ($\varphi_{\rm T}\le30^\circ$). We evaluated the method's ability to recover $\rm\Omega_p$ under various conditions. Our main results are described as follows:

\begin{enumerate}
\item For mock data sets with side-on bars ($\varphi_{\rm T}\ge50^\circ$), the method can recover the bar pattern speeds $\rm\Omega_p$ well. The model-recovered pattern speeds $\rm\Omega_p$ encompass the true pattern speeds $\rm\Omega_T$ within the model uncertainties ($1\sigma$ confidence levels, $68\%$) for 10 of 12 cases. The average model uncertainty within the $1\sigma$ confidence levels is equal to $10\%$.

\item For mock data sets with end-on bars ($\varphi_{\rm T}\le30^\circ$), if no additional priors are adopted in the modelling, the recovery accuracy of $\rm\Omega_p$ depends significantly on how much information we can obtain from the projected bars. For the case with $(\theta_{\rm T},\varphi_{\rm T})=(85^\circ,30^\circ)$, the recovery of $\rm\Omega_p$ is as good as in the side-on cases, while for the cases with $\varphi_{\rm T}=10^\circ$ where the bar is almost perpendicular to the observing plane, the model uncertainties of $\rm\Omega_p$ reach $\sim30\%$.

\item For mock data sets with end-on bars ($\varphi_{\rm T}\le30^\circ$), if the bars are forced to be strong bars with $p_{\rm bar}\le0.50$, the bar pattern speeds $\rm\Omega_p$ can be better recovered. The model-recovered pattern speeds $\rm\Omega_p$ encompass the true pattern speeds $\rm\Omega_T$ within the model uncertainties for three of four cases. The average model uncertainty within the $1\sigma$ confidence levels is equal to $14\%$, which is slightly larger than the uncertainties in the side-on cases.

\item For all the models that we create in this paper, the $2\sigma$ ($95\%$) confidence levels of $\rm\Omega_p$ always cover the true values, $\rm\Omega_T$.

\end{enumerate}

We will apply our method to edge-on barred galaxies in real observations such as GECKOS in the future. We aim to determine the bar pattern speed $\rm\Omega_p$ of these galaxies and study the possible correlations between $\rm\Omega_p$ and other physical properties.

\begin{acknowledgements} 
We have used simulations from the Auriga Project public data release \citep{Grand2024} available at \url{https://wwwmpa.mpa-garching.mpg.de/auriga/data}. The authors thank Richard J. Long for useful discussions. This work is supported by the National Science Foundation of China under Grant No. 12403017. This work is partly supported by the National Science Foundation of China (Grant No. 11821303 to SM).  
\end{acknowledgements}

\begin{appendix}
\onecolumn
\section{The MGE fitting parameters for Au-23-85-50}
In Table~\ref{table-mge-parameters}, we present the MGE fitting parameters for the surface brightness of Au-23-85-50, including the luminosity $L_j$, dispersion $\sigma_j'$, and axis ratio $q_j'$ of each Gaussian. The surface brightness fitted by MGE is calculated based on Eq.~(\ref{mge2D}).

\begin{table}[htp!]
\centering
\caption{MGE fitting parameters for Au-23-85-50.}
\begin{tabular}{|c|c|c|c|}
\hline
Component & $L_j\,[\rm 10^{10}\,L_{\odot}]$ & $\sigma_j'\,[\rm arcsec]$ & $q_j'$\\
\hline
\multirow{8}*{BP/X-shaped bar} & 1.2161   &  4.0280   &  0.3875 \\
~                              & -25.2884 &  4.7696   &  0.6976 \\
~                              & 25.8875  &  4.8152   &  0.6836 \\
~                              & -1.0308  &  4.9498   &  0.3174 \\
~                              & 3.1297   &  11.6866  &  0.3317 \\
~                              & -4.2518  &  13.0542  &  0.2380 \\
~                              & 0.2417   &  14.4589  &  0.6952 \\
~                              & 2.4328   &  16.4677  &  0.1700 \\
\hline
\multirow{3}*{Disk}            & 1.4738   &  39.5660  &  0.1700 \\
~                              & 0.4570   &  45.4114  &  0.3368 \\
~                              & 0.2433   &  54.1690  &  0.6585 \\
\hline
\end{tabular}
\tablefoot{From left to right: (1) The component to which each Gaussian belongs; (2) luminosity $L_j$; (3) dispersion $\sigma_j$; (4) axis ratio $q_j'$. The surface brightness fitted by MGE is calculated based on Eq.~(\ref{mge2D}).}
\label{table-mge-parameters}
\end{table}

\section{The influence of $\psi_{\rm bar}$ on the bar morphology}
\begin{figure*}[htp!]
    \centering
    \includegraphics[width=17.8cm]{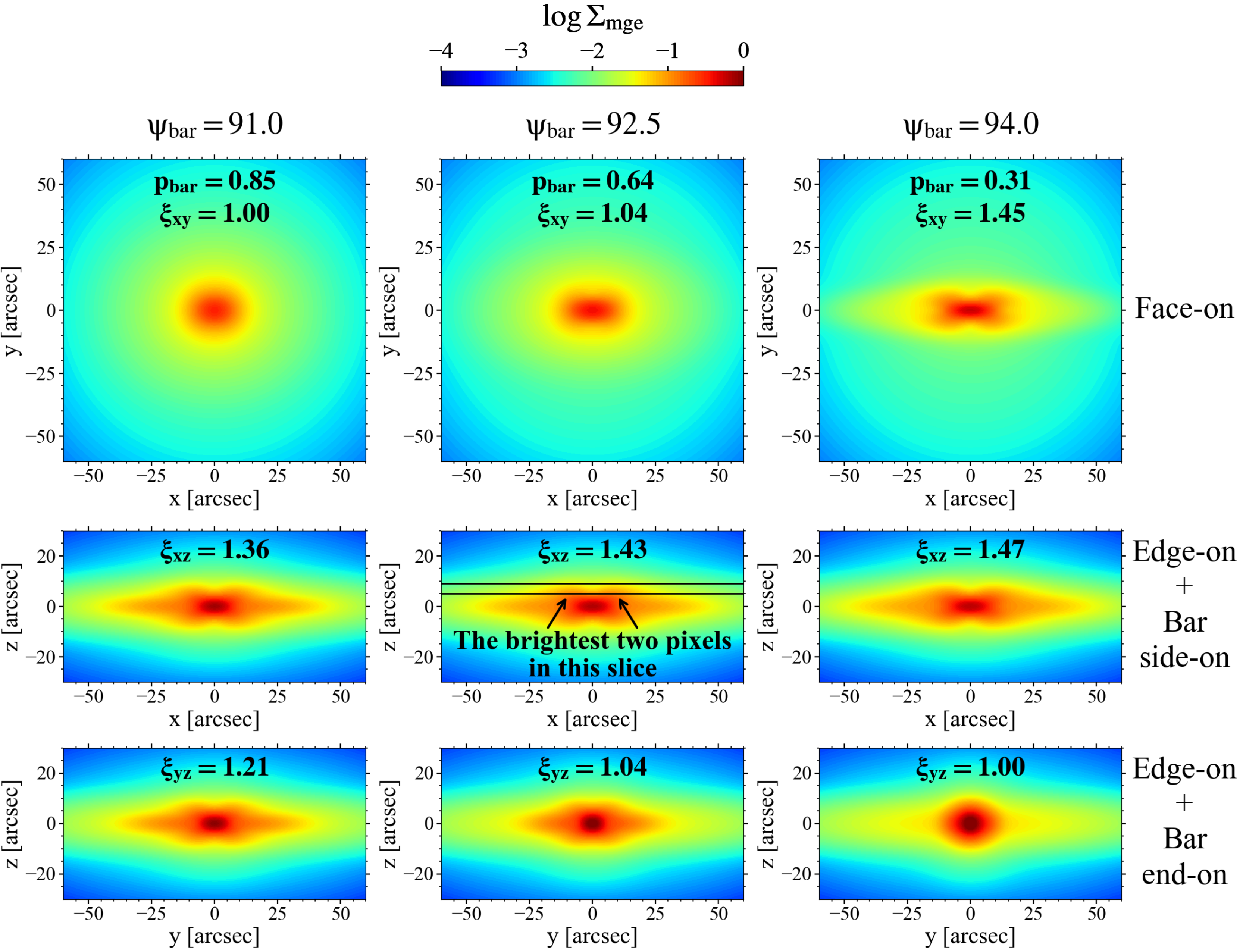}
    \caption{Variation of MGE-predicted luminosity distributions with bar position angles $\psi_{\rm bar}$. All distributions are deprojected under the true viewing angles $(\theta,\varphi)=(\theta_{\rm T},\varphi_{\rm T})=(85^\circ,50^\circ)$. From top to bottom, the panels show the normalised luminosity distributions in the intrinsic $x$-$y$, $x$-$z$, and $y$-$z$ planes. From left to right, the bar position angles are $\psi_{\rm bar}=91.0^\circ$, $92.5^\circ$, and $94.0^\circ$. The middle panels with $\psi=92.5^\circ$ correspond to our chosen distribution and are the same as the middle panels in Fig.~\ref{mge-deprojection}. The bar axis ratios $p_{\rm bar}$ in the $x$-$y$ plane and the values of the parameter $\xi$ (defined in Eqs.~\ref{Xpar1}--\ref{Xpar3}) are labelled in the figure. The lines, arrows, and labels in the central panel illustrate the calculation of $\xi_{xz}$: we first divide the $x$-$z$ plane into $\rm1\,arcsec$-wide slices along the $z$-axis; for each slice, we calculate the brightness ratio between the brightest pixel and the central pixel, and finally select the maximum ratio across all slices as $\xi_{xz}$.}
    \label{mge-deprojection-different-psi}
\end{figure*}
As described in Sect.~\ref{sec3.1.1}, minor variations in the bar position angle $\psi_{\rm bar}$ can significantly affect the intrinsic shapes of the bar when deprojecting 2D MGE to 3D. In Figure~\ref{mge-deprojection-different-psi}, we demonstrate how the MGE-predicted luminosity distribution varies with $\psi_{\rm bar}$ for Au-23-85-50 under the true viewing angles $(\theta,\varphi)=(\theta_{\rm T},\varphi_{\rm T})=(85^\circ,50^\circ)$. The middle panels with $\psi=92.5^\circ$ correspond to our chosen distribution and are the same as the middle panels in Fig.~\ref{mge-deprojection}. We also illustrate the calculation of the parameter $\xi$ as defined in Eqs.~(\ref{Xpar1})--(\ref{Xpar3}).

\section{The mass profiles for Au-23-85-50}
\begin{figure}[htp!]
    \centering
    \includegraphics[width=8.5cm]{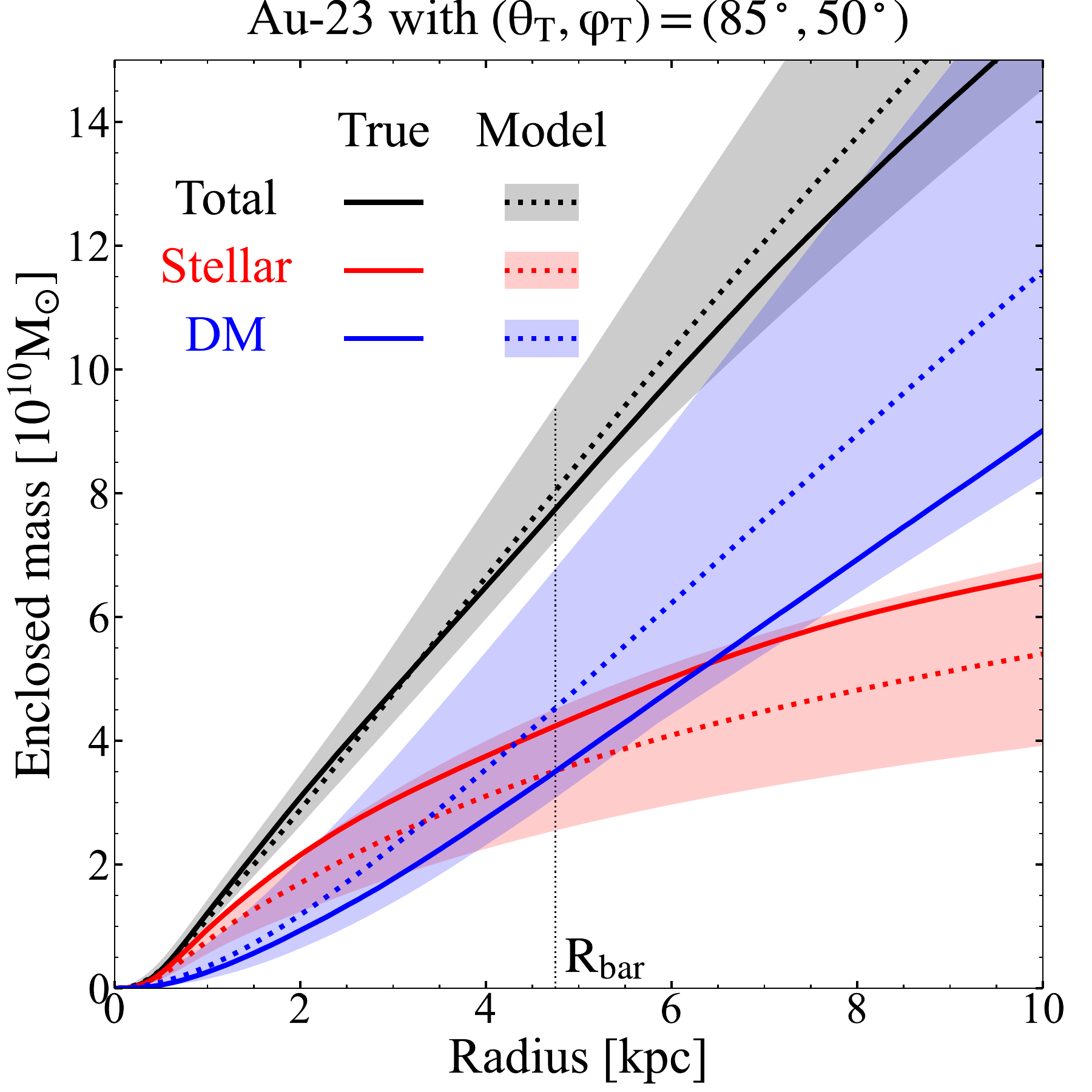}
    \caption{Enclosed mass profiles for Au-23-85-50 and comparisons with the true profiles. The black, red, and blue solid lines represent the true profiles of total mass, stellar mass, and dark matter mass, respectively. The corresponding dashed lines indicate the profiles of the best-fitting models, while the shadow regions denote the model uncertainties within $2\sigma$ confidence levels. The dotted vertical line represents the bar radius $R_{\rm bar}$ (half the bar length) from \citet{BlazquezCalero2020}.}
    \label{mass-profiles}
\end{figure}
In Fig.~\ref{mass-profiles}, we compare the enclosed mass profiles for Au-23-85-50 with the corresponding true profiles. The total mass within the bar region ($\rm\lesssim5\,kpc$) is well constrained, while large degeneracies exist between stellar mass and dark matter mass. Despite these degeneracies, the model uncertainties within $2\sigma$ confidence levels still cover the true profiles.

\section{The kinematic maps for a bar end-on case}
\begin{figure*}[htp!]
    \centering
    \includegraphics[width=16cm]{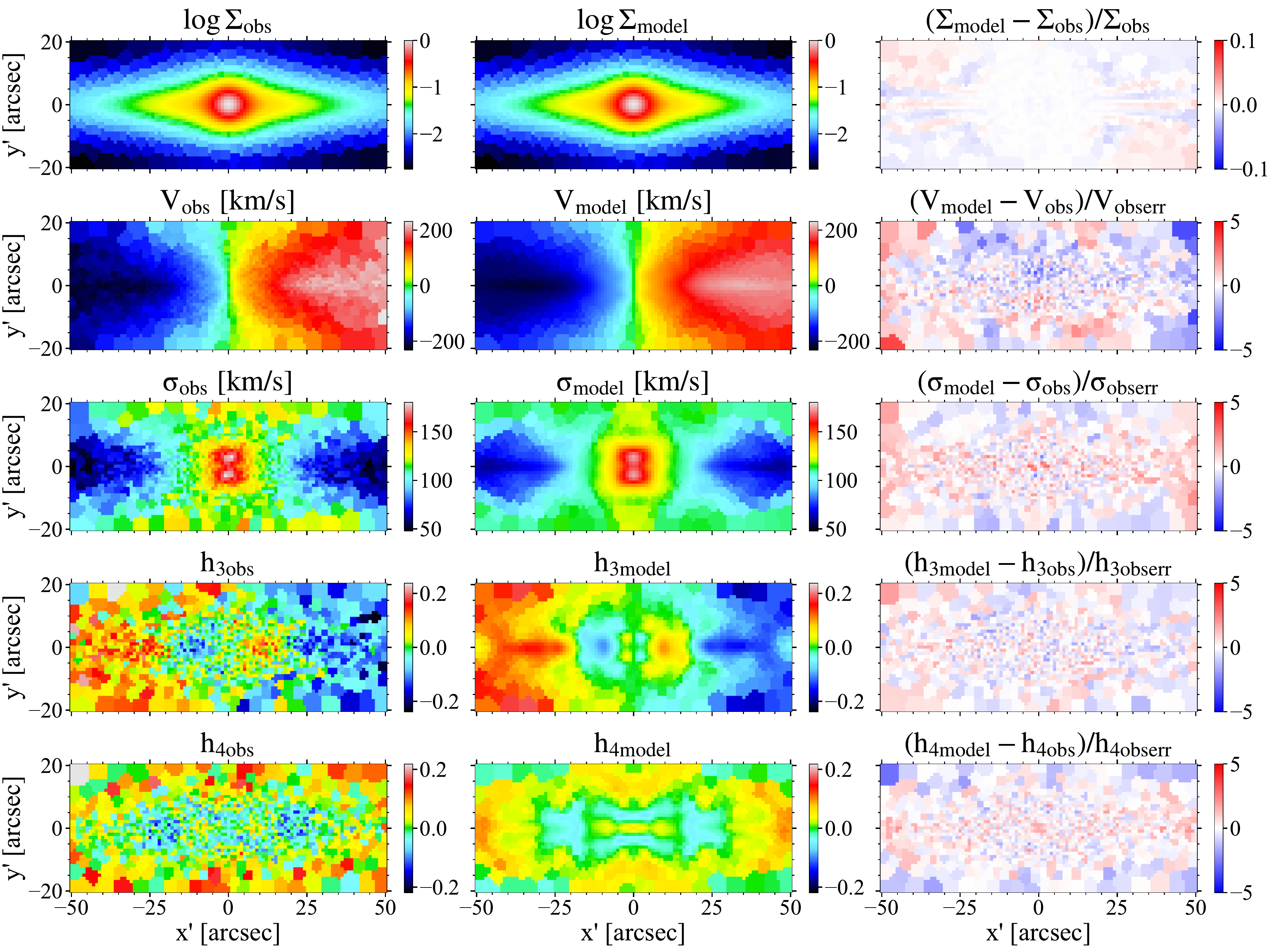}
    \caption{Similar to Fig.~\ref{best-fitting-kinematics} but for Au-23 with viewing angles $(\theta_{\rm T},\varphi_{\rm T})=(89^\circ,10^\circ)$, which corresponds to an end-on view of the bar. The maps are derived from the original models without imposing the bar to be a strong bar.}
    \label{best-fitting-kinematics-end-on}
\end{figure*}
\begin{figure*}[htp!]
    \centering
    \includegraphics[width=17.8cm]{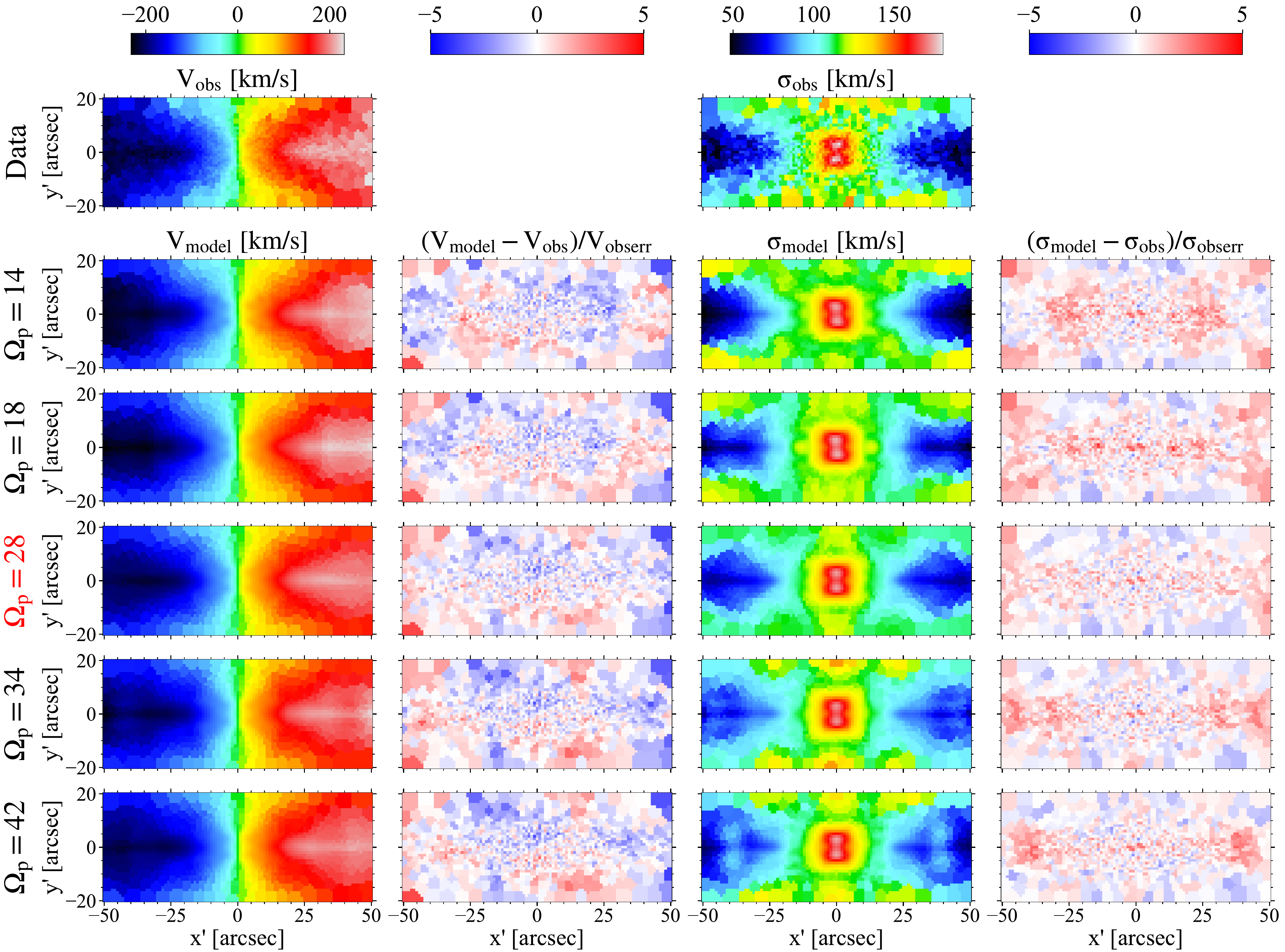}
    \caption{Similar to Fig.~\ref{kinematics-different-omega} but for Au-23 with viewing angles $(\theta_{\rm T},\varphi_{\rm T})=(89^\circ,10^\circ)$, which corresponds to an end-on view of the bar. The maps are derived from the original models without imposing the bar to be a strong bar.}
    \label{kinematics-different-omega-end-on}
\end{figure*}

In Fig.~\ref{best-fitting-kinematics-end-on}, we show the mock surface brightness, mock kinematic maps, and best-fitting model for Au-23 with $(\theta_{\rm T},\varphi_{\rm T})=(89^\circ,10^\circ)$. In Fig.~\ref{kinematics-different-omega-end-on}, we display the influence of $\rm\Omega_p$ on the kinematic maps for Au-23 with $(\theta_{\rm T},\varphi_{\rm T})=(85^\circ,50^\circ)$. The results in both figures are derived from the original models without imposing the bar to be a strong bar (see Sect.~\ref{sec5.2}).

\section{The influence of $R_0$, BP/X-shaped structures, and dark matter halos on model results}
In Fig.~\ref{chi2-vs-omega-test-R0}, we show the model-recovered bar pattern speeds for the default separating radii ($R_0=R_{0,\rm default}$), smaller ones ($R_0=0.6R_{0,\rm default}$), and larger ones ($R_0=1.3R_{0,\rm default}$). These are done for three cases: a bar side-on case Au-18-89-50, a bar side-on case Au-28-85-80, and a bar end-on case Au-23-85-30 under the strong bar assumption ($p_{\rm bar}\le0.50$). In Fig.~\ref{chi2-vs-omega-test-peanut-DMhalo}, we compare three model results: (1) the default models, which incorporate BP/X-shaped structures and allow dark matter halos to vary freely); (2) the models with constrained dark matter halos, where the differences between model-recovered and true dark matter masses within $\rm10\,kpc$ are limited to be less than $10\%$); and (3) the models ignoring BP/X-shaped structures.

\begin{figure*}[htp!]
    \centering
    \includegraphics[width=17.8cm]{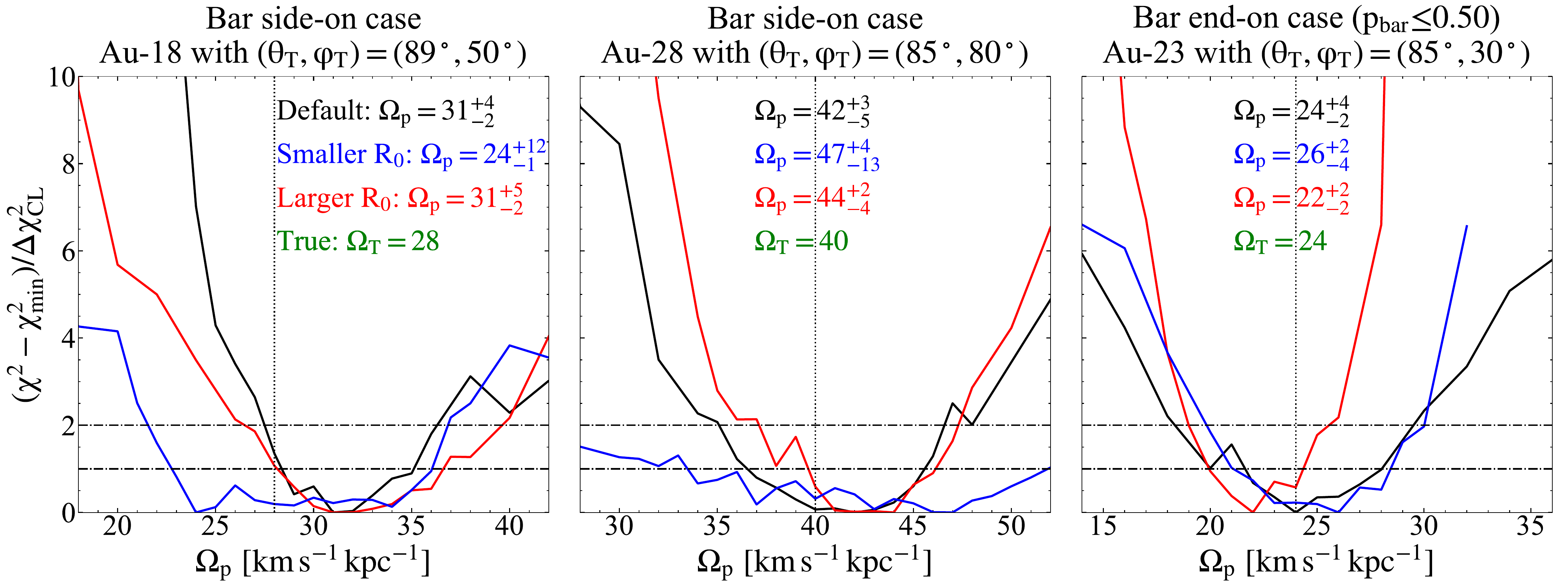}
    \caption{Influence of separating radii $R_0$ on marginalised chi-squares $\chi^2_{\rm norm}=(\chi^2-\chi^2_{\rm min})/\Delta\chi^2_{\rm CL}$ versus model-recovered bar pattern speeds $\rm\Omega_p$. From left to right: a bar side-on case Au-18-89-50, a bar side-on case Au-28-85-80, and a bar end-on case Au-23-85-30 under the strong bar assumption ($p_{\rm bar}\le0.50$). Each fold line represents a complete set of modelling runs. The black fold lines denote the model results with default separating radii ($R_0=R_{0,\rm default}$), as previously displayed in Figs.~\ref{chi2-vs-omega-side-on-cases} and~\ref{chi2-vs-omega-end-on-cases}, while the blue and red lines represent results with smaller ($R_0=0.6R_{0,\rm default}$) and larger ($R_0=1.3R_{0,\rm default}$) separating radii, respectively. The horizontal dashed lines represent $1\sigma$ ($68\%$, $\chi^2_{\rm norm}=1$) and $2\sigma$ ($95\%$, $\chi^2_{\rm norm}=2$) confidence levels. The vertical dotted lines denote the true bar pattern speeds, with their values shown by the green annotations.}
    \label{chi2-vs-omega-test-R0}
\end{figure*}
\begin{figure*}[htp!]
    \centering
    \includegraphics[width=12.2cm]{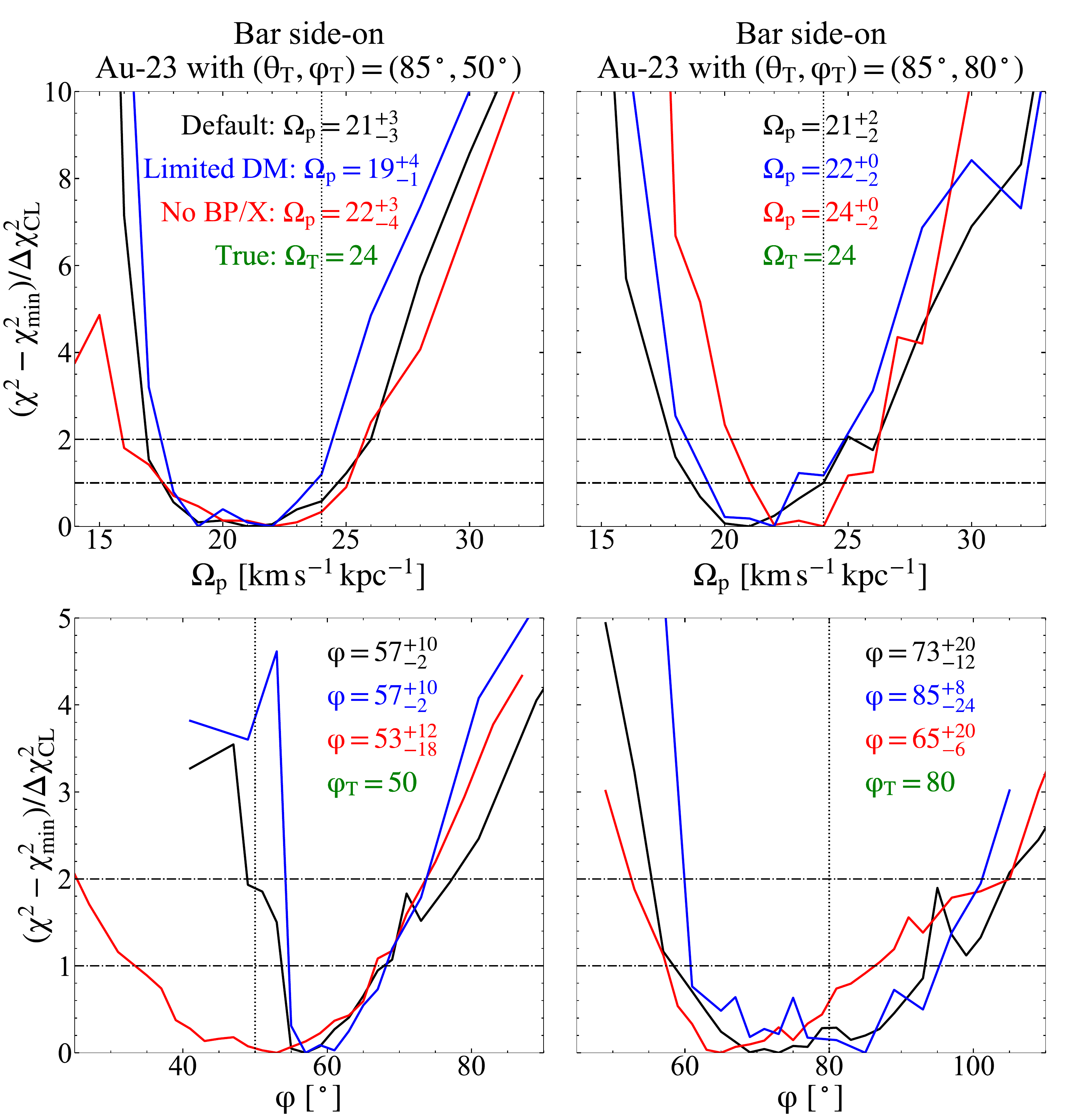}
    \caption{Influence of BP/X-shaped structures and dark matter halos on model results. The top panels show the marginalised chi-squares $\chi^2_{\rm norm}=(\chi^2-\chi^2_{\rm min})/\Delta\chi^2_{\rm CL}$ versus model-recovered bar pattern speeds $\rm\Omega_p$, while the bottom panels display $\chi^2_{\rm norm}$ versus model-recovered bar azimuthal angles $\varphi$. Left panels correspond to the results of Au-23-85-50 and right panels are for Au-23-85-80. Each fold line represents a complete set of modelling runs. The black fold lines denote the default model results, which incorporate BP/X-shaped structures and allow dark matter halos to vary freely, as previously displayed in the middle panel of Fig.~\ref{chi2-vs-omega-side-on-cases}. The blue lines correspond to results with constrained dark matter halos, requiring the differences between model-recovered and true dark matter mass within $\rm10\,kpc$ to be less than $10\%$ ($0.9M_{\rm dm,10kpc}^{\rm true}<M_{\rm dm,10kpc}^{\rm model}<1.1M_{\rm dm,10kpc}^{\rm true}$). The red lines represent the results ignoring BP/X-shaped structures. The horizontal dashed lines represent $1\sigma$ ($68\%$, $\chi^2_{\rm norm}=1$) and $2\sigma$ ($95\%$, $\chi^2_{\rm norm}=2$) confidence levels. The vertical dotted lines denote the true bar pattern speeds $\rm\Omega_T$ or azimuthal angles $\varphi_{\rm T}$, with their values shown by the green annotations.}
    \label{chi2-vs-omega-test-peanut-DMhalo}
\end{figure*}
\twocolumn
\end{appendix}
\end{nolinenumbers}
\end{document}